\documentclass[12pt]{article}
\usepackage[utf8]{inputenc}
\usepackage[british]{babel}
\usepackage{cmap}
\usepackage{lmodern}

\usepackage{amssymb, amsmath, amsthm}
\usepackage[a4paper,top=25mm,bottom=25mm,left=25mm,right=25mm]{geometry}
\usepackage{ragged2e}

\usepackage{authblk} 
\usepackage{pifont}
\usepackage{graphicx}
\usepackage[dvipsnames,svgnames,table]{xcolor}
\usepackage[figuresright]{rotating}
\usepackage{xtab} 
\usepackage{longtable} 
\usepackage{multirow}
\usepackage{footnote}
\usepackage[stable]{footmisc}
\usepackage{chngpage} 
\usepackage{pdflscape} 
\usepackage[nottoc,notlot,notlof]{tocbibind} 
\usepackage{xltabular}

\usepackage{pgfplots}
\pgfplotsset{every tick label/.append style={font=\footnotesize}}
\pgfplotsset{compat=1.14}
\usepackage{setspace}

\usepackage{array}
\newcolumntype{K}[1]{>{\centering\arraybackslash$}p{#1}<{$}}

\makesavenoteenv{tabular}
\usepackage{tabularx}
\usepackage{booktabs}
\usepackage{threeparttable}
\usepackage[referable]{threeparttablex} 
\newcolumntype{R}{>{\raggedleft\arraybackslash}X}
\newcolumntype{L}{>{\raggedright\arraybackslash}X}
\newcolumntype{C}{>{\centering\arraybackslash}X}
\newcolumntype{A}{>{\columncolor{gray!25}}C}
\newcolumntype{a}{>{\columncolor{gray!25}}c}

\newlength{\tablen}

\usepackage{dcolumn} 
\newcolumntype{.}{D{.}{.}{-1}}

\usepackage{tikz}
\usetikzlibrary{arrows, calc, matrix, patterns, positioning, trees}
\usepackage[semicolon]{natbib}
\usepackage[hyphens]{url}
\usepackage{hyperref} 
\hypersetup{
  colorlinks   = true,    		
  urlcolor     = blue,    		
  linkcolor    = blue,    		
  citecolor    = ForestGreen	
}
\usepackage{microtype}
\usepackage[justification=centering]{caption} 

\usepackage[labelformat=simple]{subcaption}

\DeclareCaptionLabelFormat{parenthesis}{(#2)}
\makeatletter
\renewcommand\p@subfigure{\arabic{figure}.}
\makeatother

\DeclareCaptionLabelFormat{parenthesis}{(#2)}
\makeatletter
\renewcommand\p@subtable{\arabic{table}.}
\makeatother

\usepackage{enumitem}

\setlist[itemize]{leftmargin=2.5\parindent}
\setlist[enumerate]{leftmargin=2.5\parindent}

\def\addlegendimage{\csname pgfplots@addlegendimage\endcsname}

\theoremstyle{plain}

\theoremstyle{definition}

\newtheorem{example}{Example}

\theoremstyle{remark}

\makeatletter
\let\@fnsymbol\@alph
\makeatother

\def\keywords{\vspace{.5em} 
{\noindent \textit{Keywords}: }}

\def\AMS{\vspace{.5em} 
{\noindent \textbf{\emph{MSC} class}: }}

\def\JEL{\vspace{.5em} 
{\noindent \textbf{\emph{JEL} classification number}: }}

\title{Refined thresholds for inconsistency: The effect of the graph associated with incomplete pairwise comparisons}
\author{Kolos Csaba \'Agoston\thanks{~Email: \emph{kolos.agoston@uni-corvinus.hu} \newline
Corvinus University of Budapest (BCE), Institute of Operations and Decision Sciences, Department of Operations Research and Actuarial Sciences, Budapest, Hungary}
$\qquad \qquad$
\href{https://sites.google.com/view/laszlocsato}{L\'aszl\'o Csat\'o}\thanks{~Corresponding author. Email: \emph{laszlo.csato@sztaki.hun-ren.hu} \newline
Institute for Computer Science and Control (SZTAKI), Hungarian Research Network (HUN-REN), Laboratory on Engineering and Management Intelligence, Research Group of Operations Research and Decision Systems, Budapest, Hungary \newline
Corvinus University of Budapest (BCE), Institute of Operations and Decision Sciences, Department of Operations Research and Actuarial Sciences, Budapest, Hungary}} 
\date{\today}

\def\Dedication{
\begin{small}
{\noindent
``\emph{The mathematical structure known as a \emph{graph} has the valuable feature of helping us to visualize, to analyze, to generalize, a situation of a problem we may encounter and, in many cases, assisting us to understand it better and possibly find a solution.}''\footnote{~Source: \citet[p.~1]{BenjaminChartrandZhang2017}.}
}
\end{small}

\flushright
\begin{small}
(Arthur Benjamin, Gary Chartrand, Ping Zhang: \emph{The Fascinating World of Graph Theory})
\end{small}

\vspace{0.5cm} 
\justify }

\begin{document}

\maketitle
\thispagestyle{empty}
\Dedication

\begin{abstract}
\noindent
The inconsistency of pairwise comparisons remains difficult to interpret in the absence of acceptability thresholds. The popular 10\% cut-off rule proposed by Saaty has recently been applied to incomplete pairwise comparison matrices, which contain some unknown comparisons. This paper refines these inconsistency thresholds: we uncover that they depend not only on the size of the matrix and the number of missing entries, but also on the undirected graph whose edges represent the known pairwise comparisons. Therefore, using our exact thresholds is especially important if the filling in patterns coincide for a large number of matrices, as has been recommended in the literature. The strong association between the new threshold values and the spectral radius of the representing graph is also demonstrated. Our results can be integrated into software to continuously monitor inconsistency during the collection of pairwise comparisons and immediately detect potential errors.


\AMS{90B50, 91B08}

\JEL{C44, D71}
\end{abstract}

\clearpage

\section{Introduction} \label{Sec1}

Pairwise comparison matrices play a central role in several multi-criteria decision-making methodologies such as the Analytic Hierarchy Process (AHP) \citep{Saaty1977, Saaty1980}. However, while pairwise comparisons efficiently decompose complex decision-making problems into simple subproblems, this comes at a price: there is no guarantee of cardinal consistency. For example, if alternative $A$ is two times better than alternative $B$, and alternative $B$ is three times better than alternative $C$, alternative $A$ is not necessarily six times better than alternative $C$, in contrast to their indirect comparison through alternative $B$. Indeed, both empirical \citep{BozokiDezsoPoeszTemesi2013} and randomly generated \citep{CsatoPetroczy2021, Csato2024a} pairwise comparison matrices are usually inconsistent. The level of inconsistency is measured by inconsistency indices \citep{Brunelli2018}.

\subsection{The need for inconsistency thresholds in the incomplete case} \label{Sec11}

The numerical value of inconsistency is difficult to interpret on its own, without a sharp threshold that separates acceptable and unacceptable levels of inconsistency. In the latter case, the original comparisons should be revised, for which several inconsistency reduction methods have been suggested \citep{MazurekPerzinaStrzakaKowalKuras2021}. For instance, \citet{BozokiFulopPoesz2015} formulate nonlinear mixed-integer optimisation problems to determine the minimal number of matrix elements to be changed in order to reduce the inconsistency below a given value.

The first inconsistency index and threshold of acceptability have been proposed by the founder of the AHP methodology \citep{Saaty1977}. It is based on dividing the inconsistency of the matrix by the average inconsistency of random pairwise comparison matrices, the so-called random index. Inconsistency can be tolerated if this ratio remains below 0.1, which directly provides a threshold for inconsistency.

Pairwise comparison matrices have a natural generalisation by allowing for missing entries \citep{Harker1987a}. Incomplete pairwise comparison matrices extend the range of potential applications to hundreds, or even millions of alternatives. For example, the pairwise comparisons of contestants in sports are usually intransitive \citep{vanOurs2024, vanOurs2025, vanOurs2026} and may be easily incomplete if some pairs of contestants do not play against each other. Such case studies are presented by \citet{BozokiCsatoTemesi2016, ChaoKouLiPeng2018, Csato2013a, PetroczyCsato2021, TemesiSzadoczkiBozoki2024}, among others. Analogously, neither completeness nor consistency holds for bilateral remittances \citep{Petroczy2021a} or students' preferences \citep{CsatoToth2020}

Even though some inconsistency indices are available for incomplete pairwise comparison matrices \citep{KulakowskiTalaga2020, SzybowskiKulakowskiPrusak2020}, determining the associated thresholds is far from trivial. Naturally, an incomplete pairwise comparison matrix can be completed \citep{AgostonCsato2024, BozokiFulopRonyai2010, CsatoAgostonBozoki2024, TekileBrunelliFedrizzi2023}, but the completion methods are usually based on \emph{minimising} inconsistency. Consequently, the thresholds derived for complete matrices become too permissive for incomplete matrices: they will accept the inconsistency of several incomplete matrices, which would be rejected unless the decision-maker provides the (almost) optimal values for all missing entries.

\subsection{The research question and our contribution} \label{Sec12}

According to our knowledge, there exists only one proposal of inconsistency thresholds for incomplete pairwise comparison matrices. \citet{AgostonCsato2022} adopt the idea of Saaty to compute the appropriate thresholds as a function of the number of alternatives and the number of missing entries. Unsurprisingly, the random index decreases if the number of unknown comparisons is higher because the inconsistency minimisation problem contains more variables, which generally implies a lower optimum.

Incomplete pairwise comparisons are often represented by undirected graphs where the vertices are the alternatives and the edges correspond to the known comparisons. The random indices, thus, the inconsistency thresholds depend on the number of vertices and edges \citep{AgostonCsato2022}. However, is it not possible that other properties of the representing graph should also be considered?

The current paper aims to address this issue.
Following the methodology of \citet{AgostonCsato2022}, we compute the average inconsistency of a large number of random incomplete pairwise comparison matrices where not only the number of vertices and edges, but the graph itself is fixed, up to isomorphism. The inconsistency threshold turns out to be strongly related to the spectral radius of the graph. Furthermore, the structure of the edges influences the proportion of randomly generated matrices that have an acceptable level of inconsistency.

\subsection{A motivating example} \label{Sec13}

Consider the following incomplete pairwise comparison matrix:
\[
\mathbf{A} = \left[
\begin{array}{K{3em} K{3em} K{3em} K{3em}}
    1     		& 2  	& \star    	& 5 \\
    1/2  	& 1       	& 4	& \star \\
    \star  		& 1/4 	& 1      	& 2 \\
    1/5  	&  \star  	& 1/2 	& 1 \\
\end{array}
\right].
\]
Matrix $\mathbf{A}$ is inconsistent as $a_{12} \cdot a_{23} \cdot a_{34} = 2 \cdot 4 \cdot 2 = 16$, but $a_{14} = 5$.
The dominant eigenvalue of the optimally completed matrix is $\lambda \approx 4.084$, resulting in an inconsistency index $\mathit{CI} \approx 0.0284$.

According to \citet[Table~2]{AgostonCsato2022}, the random index $\mathit{RI} \approx 0.306$ if there are $n = 4$ alternatives and $m = 2$ missing elements.
However, Table~\ref{Table1} reveals that the random index is $\mathit{RI} \approx 0.265$ if the positions of the two missing entries (they are placed in different columns and rows) are taken into account, too.

Therefore, the inconsistency of matrix $\mathbf{A}$ can be tolerated based on \citet{AgostonCsato2022}. However, this inconsistency is too high if the famous 10\% rule of Saaty is applied in a more sophisticated way. For pairwise comparison matrices with a graph representation analogous to the above example, the novel threshold changes the decision on 1165 matrices whose entries are drawn from the Saaty scale.

\subsection{Practical relevance} \label{Sec14}

Recently, optimal filling in patterns have been recommended for incomplete pairwise comparison matrices that provide the closest weight vectors on average to the complete case \citep{SzadoczkiBozokiTekile2022, SzadoczkiBozokiJuhaszKadenkoTsyganok2023, SzadoczkiBozoki2025}. When the pairwise comparisons are asked from the decision-makers in such a sequence, a large number of incomplete pairwise comparison matrices will have the same graph representation, see the experiment of \citet{SzadoczkiBozokiSiposGalambosi2025a}. Then it is crucial not to use the na\"ive thresholds given by \citet{AgostonCsato2022} since they assume random positions for the missing entries in the matrix. The exact thresholds provided in the current paper apply to all incomplete pairwise comparison matrices with the same graph representation; hence, the same value remains valid for matrices collected from different decision-makers, which does not complicate the analysis of inconsistency but makes it more accurate.

Our results are also important for inconsistency monitoring \citep{BozokiFulopKoczkodaj2011}. Since the pairwise comparison matrices are always filled sequentially by the decision-makers, we have a new incomplete pairwise comparison matrix after each step. By continuously checking the inconsistency of these matrices, any serious violation of inconsistency can be immediately reported to the decision-maker. Therefore, it is possible to instantly correct any error or misprint, even before all pairwise comparisons are collected. This might be especially useful due to the mental and time constraints of the experts.

\subsection{Structure} \label{Sec15}

The paper is organised as follows.
The methodology is described in Section~\ref{Sec2}. Our main findings are presented and discussed in Section~\ref{Sec3}, while Section~\ref{Sec4} offers concluding remarks.

\section{Methodology} \label{Sec2}

Inconsistency and incomplete pairwise comparison matrices are introduced in Section~\ref{Sec21}. Section~\ref{Sec22} presents the background of computing the random index, and Section~\ref{Sec23} gives an overview of our numerical analysis.

\subsection{Inconsistency and incomplete pairwise comparison matrices} \label{Sec21}

Let $n$ denote the number of alternatives.
The pairwise comparisons of the alternatives are collected into a \emph{pairwise comparison matrix} $\mathbf{A} = \left[ a_{ij} \right]$, $a_{ij} > 0$ means that the $i$th alternative is $a_{ij}$ times more preferred to the $j$th alternative. The matrix satisfies the reciprocity property: $a_{ji} = 1 / a_{ij}$ for all $1 \leq i,j \leq n$.

A pairwise comparison matrix $\mathbf{A}$ is \emph{consistent} if and only if $a_{ik} = a_{ij} \cdot a_{jk}$ holds for all $1 \leq i,j,k \leq n$. In other words, the pairwise comparisons are consistent if the direct comparison of any two alternatives coincides with their indirect comparison through any third alternative. Otherwise, the matrix is called \emph{inconsistent}.

According to the Perron--Frobenius theorem \citep{Perron1907, Frobenius1908, Frobenius1909, Frobenius1912}, a pairwise comparison matrix $\mathbf{A}$ has a dominant real eigenvalue $\lambda_{\max} \left( \mathbf{A} \right)$.
The inconsistency index of Thomas L.~Saaty \citep{Saaty1977} is:
\[
\mathit{CI} \left( \mathbf{A} \right) = \frac{\lambda_{\max} \left( \mathbf{A} \right) - n}{n-1}.
\]
$\mathit{CI} \left( \mathbf{A} \right) = 0$ if and only if the pairwise comparisons are consistent.

Saaty suggests comparing $\mathit{CI} \left( \mathbf{A} \right)$ to the \emph{random index} $\mathit{RI}$, the average $\mathit{CI}$ of a large number of pairwise comparison matrices whose entries above the diagonal are drawn independently and uniformly from the following Saaty scale:
\[
\left\{ 1/9,\, 1/8,\, 1/7, \dots ,\, 1/2,\, 1,\, 2, \dots ,\, 8,\, 9 \right\}.
\]
The values of the random index were computed several times as a function of the number of alternatives $n$ \citep{AguaronMoreno-Jimenez2003, AlonsoLamata2006, BozokiRapcsak2008, CsatoPetroczy2021, Ozdemir2005, TummalaLing1998}.

The inconsistency index divided by the random index defines the inconsistency ratio:
\[
\mathit{CR} \left( \mathbf{A} \right) = \frac{\mathit{CI} \left( \mathbf{A} \right)}{\mathit{RI}}.
\]

Incomplete pairwise comparison matrices allow for some missing comparisons. Therefore, $\mathbf{A} = \left[ a_{ij} \right]$ is an \emph{incomplete pairwise comparison matrix} if $a_{ij} > 0$ or $a_{ij} = \star$ such that $a_{ij} > 0$ implies $a_{ji} = 1 / a_{ij}$ and $a_{ij} = \star$ implies $a_{ji} = \star$.
The number of missing comparisons above the diagonal is denoted by $m$, hence, $\lvert (i,j): a_{ij} = \star \rvert = 2m$.
Incomplete pairwise comparisons have a natural graph representation.
Let $\mathbf{A}$ be an incomplete pairwise comparison matrix. The associated undirected graph is $G = (V,E)$, where the vertices correspond to the alternatives and $e_{ij} \in E$ is an edge if and only if $i \neq j$ and the pairwise comparison $a_{ij}$ is known.

The adjacency matrix $\mathbf{B} = \left[ b_{ij} \right]$ of graph $G = (V,E)$ is given by $b_{ij} = 1$ if $e_{ij} \in E$ and $b_{ij} = 0$ if $e_{ij} \notin E$.
The \emph{spectral radius} of a finite graph $G = (V,E)$ is the maximum of the absolute values of the eigenvalues of its adjacency matrix $\mathbf{B}$.

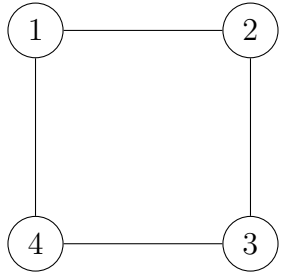
\begin{figure}[t!]
\centering
\begin{tikzpicture}[scale=1, auto=center, transform shape, >=triangle 45]
\tikzstyle{every node}=[draw,shape=circle];
  \node (n1) at (135:2) {$1$};
  \node (n2) at (45:2)  {$2$};
  \node (n3) at (315:2) {$3$};
  \node (n4) at (225:2) {$4$};

  \foreach \from/\to in {n1/n2,n1/n4,n2/n3,n3/n4}
    \draw (\from) -- (\to);
\end{tikzpicture}

\caption{The graph representation of the pairwise comparison matrix $\mathbf{A}$ in Example~\ref{Examp1}}
\label{Fig1}
\end{figure}

\begin{example} \label{Examp1}
Take the following incomplete pairwise comparison matrix:
\[
\mathbf{A} = \left[
\begin{array}{K{3em} K{3em} K{3em} K{3em}}
    1     		& 2  	& \star    	& 4 \\
    1/2  	& 1       	& 2	& \star \\
    \star  		& 1/2 	& 1      	& 2 \\
    1/4  	&  \star  	& 1/2 	& 1 \\
\end{array}
\right].
\]
Its graph representation is given in Figure~\ref{Fig1}. The spectral radius of this graph equals 2.
\end{example}

Understanding the properties of the spectral radius is an active research field in discrete mathematics \citep{Stevanovic2018}. The spectral radius of a $k$-regular graph, where each vertex is connected to the same number of vertices $k$ (has degree $k$), is equal to $k$.
Denote the maximum and minimum degree of a graph $G$ by $\Delta(G)$ and $\delta(G)$, respectively. \citet{Hong1993} asked the following question more than three decades ago: If $G$ has the smallest possible spectral radius among all graphs with $n$ vertices and $e$ edges, does $\Delta(G) - \delta(G) \leq 1$ hold, namely, is $G$ almost regular?
The answer seems to be positive, at least for dense graphs with $e \geq (n-1)(n-2)/2 - 2$ \citep{CioabaGuptaMarques2024}.
Analogously, there are recent results on upper bounds of the spectral radius \citep{GuoWangLi2019, JinZhagZhang2024}.

\subsection{The eigenvalue minimisation problem and its solution} \label{Sec22}

In order to complete an incomplete pairwise comparison matrix $\mathbf{A}$, \citet{ShiraishiObata2002} and \citet{ShiraishiObataDaigo1998} have proposed substituting the $m$ missing entries above the diagonal with positive variables collected in the vector $\mathbf{x} = \left[ x_k \right]$, $1 \leq k \leq m$. Since inconsistency ($\mathit{CI}$) is a monotone function of the dominant eigenvalue, the following optimisation problem is worth considering:
\begin{equation} \label{EV_min}
\min_{\mathbf{x} > 0} \lambda_{\max} \left( \mathbf{A} \left( \mathbf{x} \right) \right).
\end{equation}

\eqref{EV_min}~can be transformed into a convex optimisation problem \citep[Section~3]{BozokiFulopRonyai2010}. The necessary and sufficient condition for the existence of a unique solution is the connectedness of the undirected graph \citep[Theorem~2]{BozokiFulopRonyai2010}. Therefore, we will focus on incomplete pairwise comparison matrices represented by a connected graph.
\citet[Section~5]{BozokiFulopRonyai2010} presents an algorithm to obtain the optimal solution to~\eqref{EV_min}.

As discussed by \citet{AgostonCsato2022}, (at least) three possible approaches exist to calculate the random index for incomplete pairwise comparison matrices, depending on how the missing comparisons are related to the Saaty scale:
\begin{itemize}
\item
Method 1: all positive variables $x_k$ are unconstrained, the missing comparisons can be arbitrarily small or high for all $1 \leq k \leq m$;
\item
Method 2: all positive variables $x_k$ are constrained by the bounds of the Saaty scale, $1/9 \leq x_k \leq 9$ for all $1 \leq k \leq m$;
\item
Method 3: all positive variables $x_k$ are constrained directly by the Saaty scale, $x_k$ should be either an integer from 1 to 9 or its reciprocal for all $1 \leq k \leq m$.
\end{itemize}
Similar to \citet{AgostonCsato2022}, Method 2 will be implemented to avoid numerical difficulties, as well as the problems caused by using a discrete scale for the variables in~\eqref{EV_min}.

\subsection{The computation approach} \label{Sec23}

In contrast to \citet{AgostonCsato2022}, the positions of the missing entries are not chosen randomly for a given number of unknown comparisons $m$, but the associated graph is fixed in advance.
This does not make any difference if $m=1$ because all graphs with $n$ vertices and $n(n-1)/2 - 1$ edges are isomorphic. However, it can be important if $m \geq 2$.

Consequently, the random index is computed as follows:
\begin{enumerate}
\item 
We choose a connected graph $G = (V,E)$ with $n$ vertices and $m$ missing edges compared to a complete graph;
\item \label{Step2}
We generate an incomplete pairwise comparison matrix $\mathbf{A}$ with graph representation $G$ such that each element of the Saaty scale is drawn with a probability of $1/17$ for each known comparison above the diagonal;
\item
We solve the minimisation problem~\eqref{EV_min} by restricting all variables in $\mathbf{x} > 0$ between $1/9$ and $9$ (Method 2 in Section~\ref{Sec22}), which gives a complete pairwise comparison matrix $\mathbf{\hat{A}}$ as the optimal solution;
\item \label{Step4}
We compute and save the inconsistency index $\mathit{CI} \left( \mathbf{\hat{A}} \right)$;
\item
We repeat Steps~\ref{Step2}--\ref{Step4} to get 1 million random matrices with a given graph representation $G$;
\item
We compute the mean of the inconsistency indices $\mathit{CI}$ from Step~\ref{Step4}.
\end{enumerate}

\section{Results} \label{Sec3}

Section~\ref{Sec31} provides a comprehensive analysis in the case of four alternatives, and Section~\ref{Sec32} focuses on pairwise comparison matrices with two missing entries. Matrices of size five and six are discussed in Section~\ref{Sec33}. Finally, Section~\ref{Sec34} presents an approximation formula based on the number of alternatives, the number of missing elements, and the value of the spectral radius that can be used without costly computations even if $n \geq 7$.

\subsection{Incomplete matrices of size four} \label{Sec31}

If $n=4$, only the case $m=2$ unknown entries is interesting.
For $m=1$, only one graph exists since $a_{14} = \star$ can be assumed without loss of generality. For $m=3$, a connected graph is necessarily a spanning tree, and the incomplete pairwise comparison matrix has a consistent completion if the missing entries remain unconstrained.

If $m=2$, there are two possible graphs:
(1) the two missing edges are independent, they do not share any vertex (Figure~\ref{Fig1});
(2) the two missing edges have a common vertex.
The probability of case (1) is
\[
\frac{n(n-1)/2 - (n-2+n-2 + 1)}{n(n-1)/2 - 1} = 1 - \frac{4n - 8}{n(n-1) - 2},
\]
since there exist $n(n-1)/2 - 1$ different edges after one is missing, and the missing edge has $n-2$ neighbouring edges at both of its vertices. 
This value equals only $1/5 = 20\%$ for $n=4$, but converges to $1$ when $n$ increases.

The value of the random index $\mathit{RI}$ can be computed exactly: the number of possible scenarios is $17^4 = 83{,}521$ for both graphs due to the four known entries and the 17 elements of the Saaty scale. Hence, in contrast to what is done in \citet{AgostonCsato2022}, we carry out a complete enumeration instead of generating 1 million random matrices.

\begin{table}[t!]
\centering
\caption{Random indices and their implications for \\ incomplete pairwise comparison matrices of size four}
 \label{Table1}

\begin{subtable}{\textwidth}
  \centering
  \caption{Na\"ive random index $\mathit{RI}$ that is independent of the graph \citep{AgostonCsato2022}}
  \label{Table1a}
  \rowcolors{3}{}{gray!20}
    \begin{tabularx}{0.8\textwidth}{l CC} \toprule
          & Graph 1 & Graph 2 \\ \bottomrule
    Random index & 0.3061 & 0.3061 \\
    Acceptable matrices ($\mathit{CI} \leq 0.1$) & 14,789 & 12,095 \\
    Unacceptable matrices ($\mathit{CI} > 0.1$) & 68,732 & 71,426 \\
    Ratio of acceptable matrices & 17.71\% & 14.48\% \\ \bottomrule
    \end{tabularx}
\end{subtable}

\vspace{0.2cm}
\begin{subtable}{\textwidth}
  \centering
  \caption{Sophisticated random index $\mathit{RI}$ that depends on the graph (this paper)}
  \label{Table1b}
  
  \rowcolors{3}{}{gray!20}
    \begin{tabularx}{0.8\textwidth}{l CC} \toprule
          & Graph 1 & Graph 2 \\ \bottomrule
    Random index & 0.2646 & 0.3165 \\
    Acceptable matrices ($\mathit{CI} \leq 0.1$) & 13,633 & 12,343 \\
    Unacceptable matrices ($\mathit{CI} > 0.1$) & 69,888 & 71,178 \\
    Ratio of acceptable matrices & 16.32\% & 14.78\% \\ \bottomrule
    \end{tabularx}
\end{subtable}
\end{table}

Table~\ref{Table1} reports the values of the refined random indices and their implications. The structure of the graph has a non-negligible effect; $\mathit{RI}$ for Graph 1 is smaller by more than 15\% compared to $\mathit{RI}$ for Graph 2. Ignoring the graph misclassifies 1156 and 248 incomplete pairwise comparison matrices in the two possible cases, respectively. Even though the standard $\mathit{CI}$ threshold of 0.1 becomes more (less) restrictive for Graph 1 (Graph 2) if the structure of the graph is taken into account, the number of matrices with an acceptable inconsistency is still higher for Graph 2 than for Graph 1.

Considering all possible matrices allows us to exactly compute the difference between Methods 1--3 mentioned in Section~\ref{Sec22}, too.
If the missing entries are continuous and not constrained to the interval $\left[ 1/9; \, 9 \right]$ (Method~1), a closed formula exists for the optimal completion if $n=4$: \citet{CsatoAgostonBozoki2024} prove that the incomplete eigenvector and the incomplete logarithmic least squares methods \citep{Kwiesielewicz1996, BozokiFulopRonyai2010, BozokiTsyganok2019} imply the same optimal completion, and the optimal completion according to the incomplete logarithmic least squares method can be obtained as a solution of a linear system of equations \citep{KaiserSerlin1978, BozokiFulopRonyai2010}. Thus, there is no need to use the algorithm suggested by \citet[Section~5]{BozokiFulopRonyai2010} to solve optimisation problem~\eqref{EV_min}. The random indices are 0.2528 and 0.2923 for Graphs 1 and 2, respectively, which are reduced by 7.6\% and 4.4\% compared to the values presented in Table~\ref{Table1}.

If the missing entries are chosen from the discrete Saaty scale (Method~3), the random indices are 0.2663 and 0.3174 for Graphs 1 and 2, respectively, which are higher by 0.31\% and 0.65\% compared to the values presented in Table~\ref{Table1}. Therefore, although our continuous relaxation clearly lowers the random index, the extent of this decrease compared to using the discrete scale seems to be minimal.

\subsection{Incomplete matrices with two missing elements} \label{Sec32}

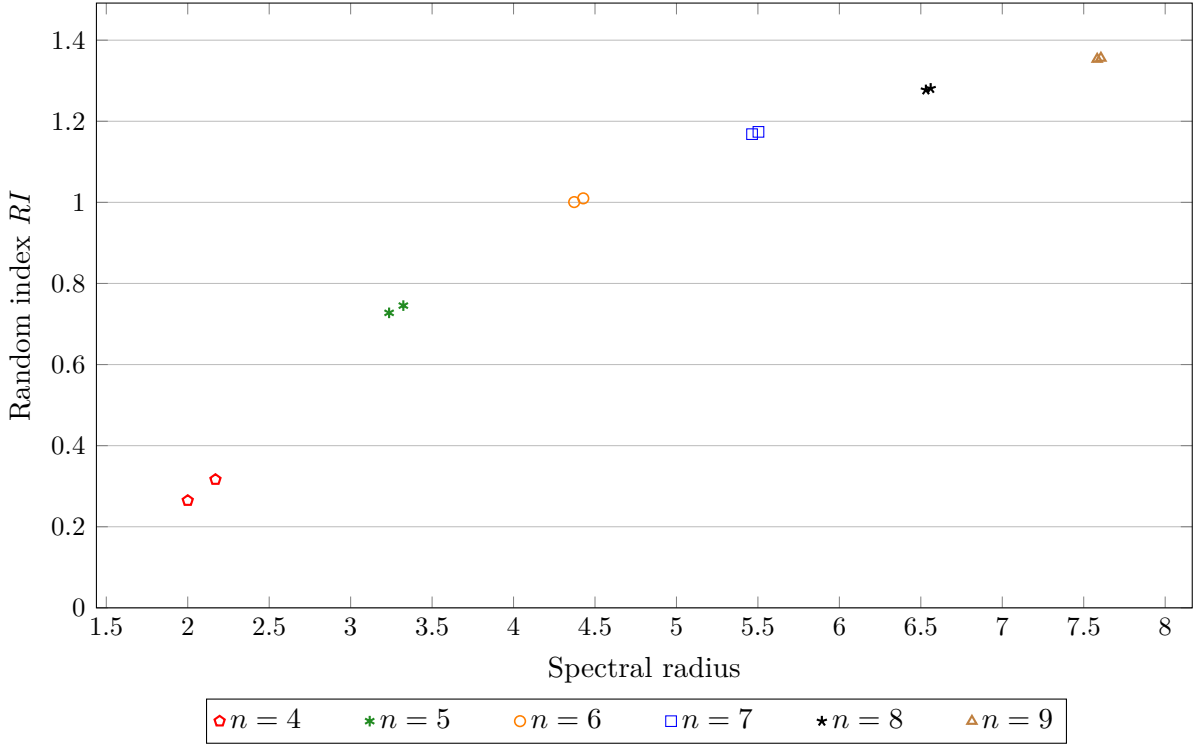
\begin{figure}[t!]
\centering

\begin{tikzpicture}
\begin{axis}[
xlabel = Spectral radius,
x label style = {font=\small},
ylabel = Random index $\mathit {RI}$,
y label style = {font=\small},
width = \textwidth,
height = 0.6\textwidth,
ymajorgrids = true,
ymin = 0,
legend style = {font=\small,at={(0.1,-0.15)},anchor=north west,legend columns=6},
legend entries = {$n = 4 \qquad$,$n = 5 \qquad$,$n = 6 \qquad$,$n = 7 \qquad$,$n = 8 \qquad$,$n = 9$}
] 
\addplot [red, thick, only marks, mark=pentagon, mark size=2pt, mark options=solid] coordinates {
(2,0.2645727)
(2.1700865,0.3164516)
};
\addplot [ForestGreen, thick, only marks, mark=asterisk, mark size=2pt, mark options=solid] coordinates {
(3.236068,0.7275285)
(3.3234043,0.7453902)
};
\addplot [orange, thick, only marks, mark=o, mark size=2pt, mark options={solid,semithick}] coordinates {
(4.372281,1.000693)
(4.4278789,1.009885)
};
\addplot [blue, thick, only marks, mark=square, mark size=2pt, mark options={solid,thin}] coordinates {
(5.464102,1.168245)
(5.5033076,1.173869)
};
\addplot [black, thick, only marks, mark=star, mark size=2pt, mark options=solid] coordinates {
(6.531129,1.277332)
(6.5605253,1.281065)
};
\addplot [brown, thick, only marks, mark=triangle, mark size=2pt, mark options=solid] coordinates {
(7.582576,1.353207)
(7.6055513,1.355913)
};
\end{axis}
\end{tikzpicture}

\captionsetup{justification=centering}
\caption{Spectral radius and random index for incomplete \\ pairwise comparison matrices with $m=2$ missing entries}
\label{Fig2}

\end{figure}


Figure~\ref{Fig2} shows the random indices for the two different incomplete pairwise comparison matrices with two missing entries as a function of the spectral radius of the representing graph if the number of alternatives is between four and nine. The plot has two important messages. First, a higher spectral radius is always associated with a higher value of $\mathit{RI}$ if $m=2$. Second, the difference between the two cases gradually disappears when the number of alternatives grows, that is, the random indices provided in \citet{AgostonCsato2022} become less distorted. This is not surprising since fixing the number of unknown comparisons means that the ratio of known elements increases together with $n$.

\subsection{Incomplete matrices of size five and six} \label{Sec33}

If $n=5$, all cases are investigated, that is, the number of missing entries varies from 1 to 5 because $m=6$ implies the existence of a consistent completion (if the unknown entries remain unconstrained). Only one graph exists if $m=1$, when the threshold given in \citet{AgostonCsato2022} remains valid.

\begin{figure}[t!]
\centering
\begin{subfigure}{0.24\textwidth}
\centering
\caption{$G_{2,1}: \left[ 4,4,3,3,2 \right]$}
\label{Fig3_1}

\begin{tikzpicture}[scale=0.85, auto=center, transform shape, >=triangle 45]
\tikzstyle{every node}=[draw,shape=circle];
  \node (n1) at (216:2) {};
  \node (n2) at (144:2) {};
  \node (n3) at (72:2)  {};
  \node (n4) at (0:2)   {};
  \node (n5) at (288:2) {};

  \foreach \from/\to in {n1/n4,n1/n5,n2/n3,n2/n4,n2/n5,n3/n4,n3/n5,n4/n5}
    \draw[dotted] (\from) -- (\to);
  \foreach \from/\to in {n1/n2,n1/n3}
    \draw[red,thick] (\from) -- (\to);
\end{tikzpicture}
\end{subfigure}
\begin{subfigure}{0.24\textwidth}
\centering
\caption{$G_{2,2}: \left[ 4,3,3,3,3 \right]$}
\label{Fig3_2}

\begin{tikzpicture}[scale=0.85, auto=center, transform shape, >=triangle 45]
\tikzstyle{every node}=[draw,shape=circle];
  \node (n1) at (216:2) {};
  \node (n2) at (144:2) {};
  \node (n3) at (72:2)  {};
  \node (n4) at (0:2)   {};
  \node (n5) at (288:2) {};

  \foreach \from/\to in {n1/n3,n1/n4,n1/n5,n2/n3,n2/n4,n2/n5,n3/n5,n4/n5}
    \draw[dotted] (\from) -- (\to);
  \foreach \from/\to in {n1/n2,n3/n4}
    \draw[red,thick] (\from) -- (\to);
\end{tikzpicture}
\end{subfigure}
\begin{subfigure}{0.24\textwidth}
\centering
\caption{$G_{3,1}: \left[ 4,3,3,3,1 \right]$}
\label{Fig3_3}

\begin{tikzpicture}[scale=0.85, auto=center, transform shape, >=triangle 45]
\tikzstyle{every node}=[draw,shape=circle];
  \node (n1) at (216:2) {};
  \node (n2) at (144:2) {};
  \node (n3) at (72:2)  {};
  \node (n4) at (0:2)   {};
  \node (n5) at (288:2) {};

  \foreach \from/\to in {n1/n3,n1/n4,n1/n5,n2/n3,n2/n4,n2/n5,n3/n4,n3/n5,n4/n5}
    \draw[dotted] (\from) -- (\to);
  \foreach \from/\to in {n1/n2,n1/n3,n1/n4}
    \draw[red,thick] (\from) -- (\to);
\end{tikzpicture}
\end{subfigure}
\begin{subfigure}{0.24\textwidth}
\centering
\caption{$G_{3,2}: \left[ 4,3,3,2,2 \right]$}
\label{Fig3_4}

\begin{tikzpicture}[scale=0.85, auto=center, transform shape, >=triangle 45]
\tikzstyle{every node}=[draw,shape=circle];
  \node (n1) at (216:2) {};
  \node (n2) at (144:2) {};
  \node (n3) at (72:2)  {};
  \node (n4) at (0:2)   {};
  \node (n5) at (288:2) {};

  \foreach \from/\to in {n1/n3,n1/n4,n1/n5,n2/n3,n2/n4,n2/n5,n3/n4,n3/n5,n4/n5}
    \draw[dotted] (\from) -- (\to);
  \foreach \from/\to in {n1/n2,n2/n3,n3/n4}
    \draw[red,thick] (\from) -- (\to);
\end{tikzpicture}
\end{subfigure}

\vspace{0.2cm}
\begin{subfigure}{0.24\textwidth}
\centering
\caption{$G_{3,3}: \left[ 3,3,3,3,2 \right]$}
\label{Fig3_5}

\begin{tikzpicture}[scale=0.85, auto=center, transform shape, >=triangle 45]
\tikzstyle{every node}=[draw,shape=circle];
  \node (n1) at (216:2) {};
  \node (n2) at (144:2) {};
  \node (n3) at (72:2)  {};
  \node (n4) at (0:2)   {};
  \node (n5) at (288:2) {};

  \foreach \from/\to in {n1/n3,n1/n4,n1/n5,n2/n3,n2/n4,n2/n5,n3/n4,n3/n5,n4/n5}
    \draw[dotted] (\from) -- (\to);
  \foreach \from/\to in {n1/n2,n1/n3,n4/n5}
    \draw[red,thick] (\from) -- (\to);
\end{tikzpicture}
\end{subfigure}
\begin{subfigure}{0.24\textwidth}
\centering
\caption{$G_{3,4}: \left[ 4,4,2,2,2 \right]$}
\label{Fig3_6}

\begin{tikzpicture}[scale=0.85, auto=center, transform shape, >=triangle 45]
\tikzstyle{every node}=[draw,shape=circle];
  \node (n1) at (216:2) {};
  \node (n2) at (144:2) {};
  \node (n3) at (72:2)  {};
  \node (n4) at (0:2)   {};
  \node (n5) at (288:2) {};

  \foreach \from/\to in {n1/n3,n1/n4,n1/n5,n2/n3,n2/n4,n2/n5,n3/n4,n3/n5,n4/n5}
    \draw[dotted] (\from) -- (\to);
  \foreach \from/\to in {n1/n2,n1/n3,n2/n3}
    \draw[red,thick] (\from) -- (\to);
\end{tikzpicture}
\end{subfigure}
\begin{subfigure}{0.24\textwidth}
\centering
\caption{$G_{4,1}: \left[ 3,3,2,2,2 \right]$}
\label{Fig3_7}

\begin{tikzpicture}[scale=0.85, auto=center, transform shape, >=triangle 45]
\tikzstyle{every node}=[draw,shape=circle];
  \node (n1) at (216:2) {};
  \node (n2) at (144:2) {};
  \node (n3) at (72:2)  {};
  \node (n4) at (0:2)   {};
  \node (n5) at (288:2) {};

  \foreach \from/\to in {n1/n3,n1/n4,n1/n5,n2/n3,n2/n4,n2/n5,n3/n4,n3/n5,n4/n5}
    \draw[dotted] (\from) -- (\to);
  \foreach \from/\to in {n1/n2,n1/n3,n2/n3,n4/n5}
    \draw[red,thick] (\from) -- (\to);
\end{tikzpicture}
\end{subfigure}
\begin{subfigure}{0.24\textwidth}
\centering
\caption{$G_{4,2}: \left[ 3,3,2,2,2 \right]$}
\label{Fig3_8}

\begin{tikzpicture}[scale=0.85, auto=center, transform shape, >=triangle 45]
\tikzstyle{every node}=[draw,shape=circle];
  \node (n1) at (216:2) {};
  \node (n2) at (144:2) {};
  \node (n3) at (72:2)  {};
  \node (n4) at (0:2)   {};
  \node (n5) at (288:2) {};

  \foreach \from/\to in {n1/n3,n1/n4,n1/n5,n2/n3,n2/n4,n2/n5,n3/n4,n3/n5,n4/n5}
    \draw[dotted] (\from) -- (\to);
  \foreach \from/\to in {n1/n2,n2/n3,n3/n4,n4/n5}
    \draw[red,thick] (\from) -- (\to);
\end{tikzpicture}
\end{subfigure}

\vspace{0.2cm}
\begin{subfigure}{0.24\textwidth}
\centering
\caption{$G_{4,3}: \left[ 4,2,2,2,2 \right]$}
\label{Fig1_9}

\begin{tikzpicture}[scale=0.85, auto=center, transform shape, >=triangle 45]
\tikzstyle{every node}=[draw,shape=circle];
  \node (n1) at (216:2) {};
  \node (n2) at (144:2) {};
  \node (n3) at (72:2)  {};
  \node (n4) at (0:2)   {};
  \node (n5) at (288:2) {};

  \foreach \from/\to in {n1/n3,n1/n4,n1/n5,n2/n3,n2/n4,n2/n5,n3/n4,n3/n5,n4/n5}
    \draw[dotted] (\from) -- (\to);
  \foreach \from/\to in {n1/n2,n2/n3,n3/n4,n1/n4}
    \draw[red,thick] (\from) -- (\to);
\end{tikzpicture}
\end{subfigure}
\begin{subfigure}{0.24\textwidth}
\centering
\caption{$G_{4,4}: \left[ 4,3,2,2,1 \right]$}
\label{Fig1_10}

\begin{tikzpicture}[scale=0.85, auto=center, transform shape, >=triangle 45]
\tikzstyle{every node}=[draw,shape=circle];
  \node (n1) at (216:2) {};
  \node (n2) at (144:2) {};
  \node (n3) at (72:2)  {};
  \node (n4) at (0:2)   {};
  \node (n5) at (288:2) {};

  \foreach \from/\to in {n1/n3,n1/n4,n1/n5,n2/n3,n2/n4,n2/n5,n3/n4,n3/n5,n4/n5}
    \draw[dotted] (\from) -- (\to);
  \foreach \from/\to in {n1/n2,n1/n3,n2/n3,n1/n4}
    \draw[red,thick] (\from) -- (\to);
\end{tikzpicture}
\end{subfigure}
\begin{subfigure}{0.24\textwidth}
\centering
\caption{$G_{4,5}: \left[ 3,3,3,2,1 \right]$}
\label{Fig1_11}

\begin{tikzpicture}[scale=0.85, auto=center, transform shape, >=triangle 45]
\tikzstyle{every node}=[draw,shape=circle];
  \node (n1) at (216:2) {};
  \node (n2) at (144:2) {};
  \node (n3) at (72:2)  {};
  \node (n4) at (0:2)   {};
  \node (n5) at (288:2) {};

  \foreach \from/\to in {n1/n3,n1/n4,n1/n5,n2/n3,n2/n4,n2/n5,n3/n4,n3/n5,n4/n5}
    \draw[dotted] (\from) -- (\to);
  \foreach \from/\to in {n1/n2,n1/n3,n1/n4,n4/n5}
    \draw[red,thick] (\from) -- (\to);
\end{tikzpicture}
\end{subfigure}
\begin{subfigure}{0.24\textwidth}
\centering
\caption{$G_{5,1}: \left[ 4,2,2,1,1 \right]$}
\label{Fig1_12}

\begin{tikzpicture}[scale=0.85, auto=center, transform shape, >=triangle 45]
\tikzstyle{every node}=[draw,shape=circle];
  \node (n1) at (216:2) {};
  \node (n2) at (144:2) {};
  \node (n3) at (72:2)  {};
  \node (n4) at (0:2)   {};
  \node (n5) at (288:2) {};

  \foreach \from/\to in {n1/n3,n1/n4,n1/n5,n2/n3,n2/n4,n2/n5,n3/n4,n3/n5,n4/n5}
    \draw[dotted] (\from) -- (\to);
  \foreach \from/\to in {n1/n2,n1/n3,n1/n4,n2/n3,n3/n4}
    \draw[red,thick] (\from) -- (\to);
\end{tikzpicture}
\end{subfigure}

\vspace{0.2cm}
\begin{subfigure}{0.24\textwidth}
\centering
\caption{$G_{5,2}: \left[ 3,3,2,1,1 \right]$}
\label{Fig1_13}

\begin{tikzpicture}[scale=0.85, auto=center, transform shape, >=triangle 45]
\tikzstyle{every node}=[draw,shape=circle];
  \node (n1) at (216:2) {};
  \node (n2) at (144:2) {};
  \node (n3) at (72:2)  {};
  \node (n4) at (0:2)   {};
  \node (n5) at (288:2) {};

  \foreach \from/\to in {n1/n3,n1/n4,n1/n5,n2/n3,n2/n4,n2/n5,n3/n4,n3/n5,n4/n5}
    \draw[dotted] (\from) -- (\to);
  \foreach \from/\to in {n1/n2,n1/n3,n2/n3,n1/n5,n3/n4}
    \draw[red,thick] (\from) -- (\to);
\end{tikzpicture}
\end{subfigure}
\begin{subfigure}{0.24\textwidth}
\centering
\caption{$G_{5,3}: \left[ 3,2,2,2,1 \right]$}
\label{Fig1_14}

\begin{tikzpicture}[scale=0.85, auto=center, transform shape, >=triangle 45]
\tikzstyle{every node}=[draw,shape=circle];
  \node (n1) at (216:2) {};
  \node (n2) at (144:2) {};
  \node (n3) at (72:2)  {};
  \node (n4) at (0:2)   {};
  \node (n5) at (288:2) {};

  \foreach \from/\to in {n1/n3,n1/n4,n1/n5,n2/n3,n2/n4,n2/n5,n3/n4,n3/n5,n4/n5}
    \draw[dotted] (\from) -- (\to);
  \foreach \from/\to in {n1/n2,n1/n3,n2/n3,n1/n4,n4/n5}
    \draw[red,thick] (\from) -- (\to);
\end{tikzpicture}
\end{subfigure}
\begin{subfigure}{0.24\textwidth}
\centering
\caption{$G_{5,4}: \left[ 3,2,2,2,1 \right]$}
\label{Fig1_15}

\begin{tikzpicture}[scale=0.85, auto=center, transform shape, >=triangle 45]
\tikzstyle{every node}=[draw,shape=circle];
  \node (n1) at (216:2) {};
  \node (n2) at (144:2) {};
  \node (n3) at (72:2)  {};
  \node (n4) at (0:2)   {};
  \node (n5) at (288:2) {};

  \foreach \from/\to in {n1/n3,n1/n4,n1/n5,n2/n3,n2/n4,n2/n5,n3/n4,n3/n5,n4/n5}
    \draw[dotted] (\from) -- (\to);
  \foreach \from/\to in {n1/n2,n2/n3,n3/n4,n1/n4,n4/n5}
    \draw[red,thick] (\from) -- (\to);
\end{tikzpicture}
\end{subfigure}
\begin{subfigure}{0.24\textwidth}
\centering
\caption{$G_{5,5}: \left[ 2,2,2,2,2 \right]$}
\label{Fig1_16}

\begin{tikzpicture}[scale=0.85, auto=center, transform shape, >=triangle 45]
\tikzstyle{every node}=[draw,shape=circle];
  \node (n1) at (216:2) {};
  \node (n2) at (144:2) {};
  \node (n3) at (72:2)  {};
  \node (n4) at (0:2)   {};
  \node (n5) at (288:2) {};

  \foreach \from/\to in {n1/n3,n1/n4,n1/n5,n2/n3,n2/n4,n2/n5,n3/n4,n3/n5,n4/n5}
    \draw[dotted] (\from) -- (\to);
  \foreach \from/\to in {n1/n2,n2/n3,n3/n4,n4/n5,n1/n5}
    \draw[red,thick] (\from) -- (\to);
\end{tikzpicture}
\end{subfigure}

\captionsetup{justification=centerfirst}
\caption{Possible graph representations of incomplete pairwise comparison matrices of size five with at least two and at most five missing entries \\ \vspace{0.2cm}
\footnotesize{\emph{Notes}: Dotted lines indicate the known comparisons, thick red lines indicate the unknown comparisons. \\
Graph $G_{i,j}$ has $i$ missing edges. The vectors show the degree distribution of the associated graph. \\
See Table~\ref{Table_A1} for the spectral radii and random indices of these incomplete pairwise comparison matrices.}}
\label{Fig3}
\end{figure}
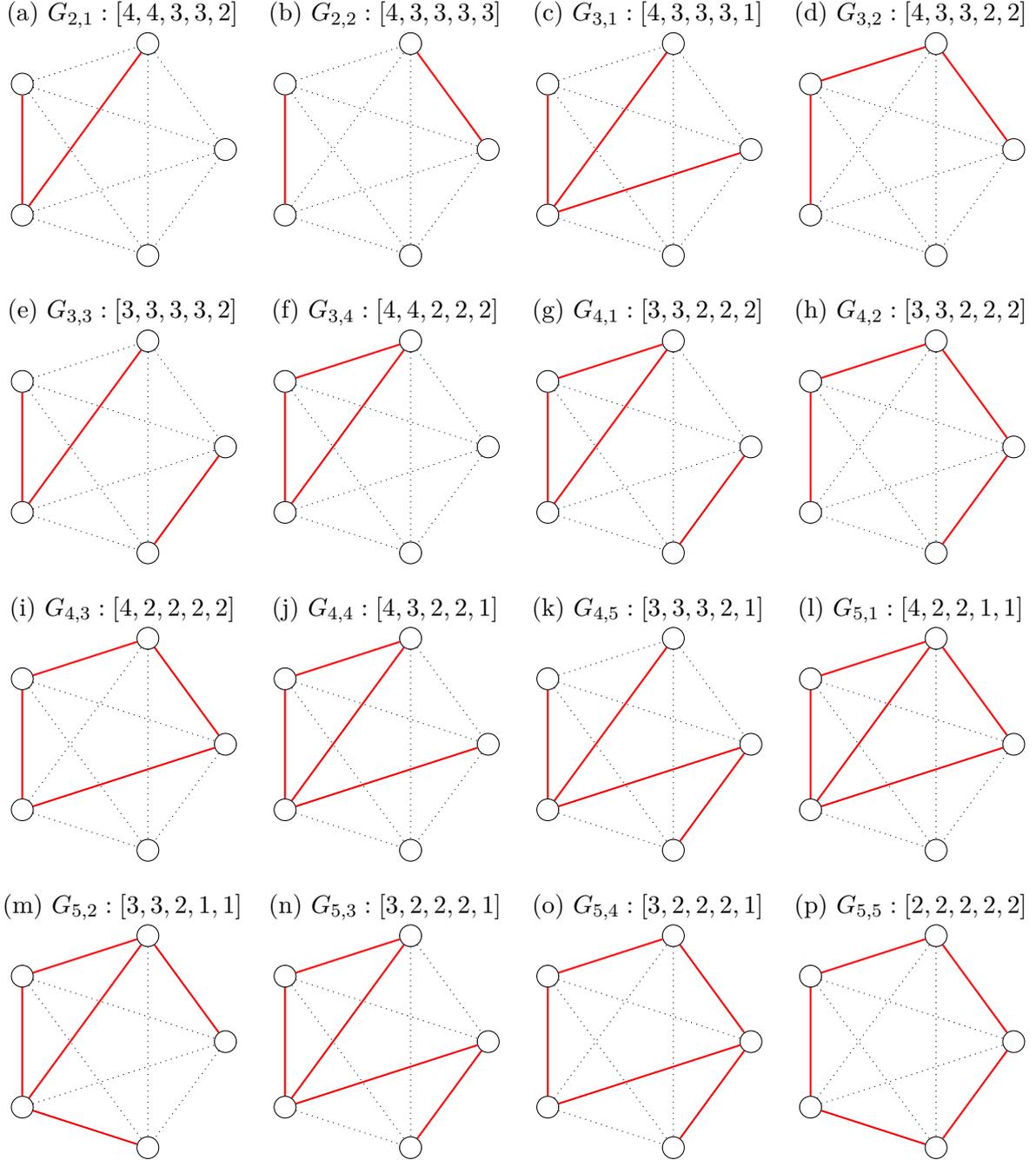


Figure~\ref{Fig3} presents the 16 possible graphs if $n=5$ and $2 \leq m \leq 5$.
The $1 + 2 + 4 + 5 + 5 = 17$ random indices---depending on the structure of the graph---are reported in Table~\ref{Table_A1} in the Appendix, together with their probability of occurrence if the number of missing entries $m$ is fixed. The probabilities are estimated by simulations, contrary to the exact values derived in Section~\ref{Sec31}. For a fixed value of $m$, the spectral radius is the smallest if the graph is (almost) regular ($G_{2,2}$, $G_{3,3}$, $G_{4,1}$, $G_{5,5}$). These graphs have the smallest probability to occur except for $G_{3,3}$, and the corresponding random index $\mathit{RI}$ is lower than for any other graphs having the same number of edges.



\begin{figure}[t!]
\centering
\begin{subfigure}{0.24\textwidth}
\centering
\caption{$G_1: \left[ 4,4,3,3,2,2 \right]$}
\label{Fig4_1}

\begin{tikzpicture}[scale=0.85, auto=center, transform shape, >=triangle 45]
\tikzstyle{every node}=[draw,shape=circle];
  \node (n1) at (180:2) {};
  \node (n2) at (120:2) {};
  \node (n3) at (60:2)  {};
  \node (n4) at (0:2)   {};
  \node (n5) at (300:2) {};
  \node (n6) at (240:2) {};

  \foreach \from/\to in {n1/n3,n1/n4,n1/n5,n1/n6,n2/n4,n2/n6,n3/n5,n3/n6,n4/n6}
    \draw[dotted] (\from) -- (\to);
  \foreach \from/\to in {n1/n2,n2/n3,n2/n5,n3/n4,n4/n5,n5/n6}
    \draw[red,thick] (\from) -- (\to);
\end{tikzpicture}
\end{subfigure}
\begin{subfigure}{0.24\textwidth}
\centering
\caption{$G_2: \left[ 4,4,3,3,2,2 \right]$}
\label{Fig4_2}

\begin{tikzpicture}[scale=0.85, auto=center, transform shape, >=triangle 45]
\tikzstyle{every node}=[draw,shape=circle];
  \node (n1) at (180:2) {};
  \node (n2) at (120:2) {};
  \node (n3) at (60:2)  {};
  \node (n4) at (0:2)   {};
  \node (n5) at (300:2) {};
  \node (n6) at (240:2) {};

  \foreach \from/\to in {n1/n3,n1/n4,n1/n5,n1/n6,n2/n4,n2/n5,n3/n4,n3/n6,n4/n6}
    \draw[dotted] (\from) -- (\to);
  \foreach \from/\to in {n1/n2,n2/n3,n2/n6,n3/n5,n4/n5,n5/n6}
    \draw[red,thick] (\from) -- (\to);
\end{tikzpicture}
\end{subfigure}
\begin{subfigure}{0.24\textwidth}
\centering
\caption{$G_3: \left[ 4,3,3,3,3,2 \right]$}
\label{Fig4_3}

\begin{tikzpicture}[scale=0.85, auto=center, transform shape, >=triangle 45]
\tikzstyle{every node}=[draw,shape=circle];
  \node (n1) at (180:2) {};
  \node (n2) at (120:2) {};
  \node (n3) at (60:2)  {};
  \node (n4) at (0:2)   {};
  \node (n5) at (300:2) {};
  \node (n6) at (240:2) {};

  \foreach \from/\to in {n1/n2,n1/n3,n1/n4,n1/n5,n2/n4,n2/n6,n3/n5,n3/n6,n4/n6}
    \draw[dotted] (\from) -- (\to);
  \foreach \from/\to in {n1/n6,n2/n3,n2/n5,n3/n4,n4/n5,n5/n6}
    \draw[red,thick] (\from) -- (\to);
\end{tikzpicture}
\end{subfigure}
\begin{subfigure}{0.24\textwidth}
\centering
\caption{$G_4: \left[ 5,4,3,2,2,2 \right]$}
\label{Fig4_4}

\begin{tikzpicture}[scale=0.85, auto=center, transform shape, >=triangle 45]
\tikzstyle{every node}=[draw,shape=circle];
  \node (n1) at (180:2) {};
  \node (n2) at (120:2) {};
  \node (n3) at (60:2)  {};
  \node (n4) at (0:2)   {};
  \node (n5) at (300:2) {};
  \node (n6) at (240:2) {};

  \foreach \from/\to in {n1/n2,n1/n3,n1/n4,n1/n5,n1/n6,n2/n4,n3/n6,n4/n6,n5/n6}
    \draw[dotted] (\from) -- (\to);
  \foreach \from/\to in {n2/n3,n2/n5,n2/n6,n3/n4,n3/n5,n4/n5}
    \draw[red,thick] (\from) -- (\to);
\end{tikzpicture}
\end{subfigure}

\vspace{0.17cm}
\begin{subfigure}{0.24\textwidth}
\centering
\caption{$G_5: \left[ 4,4,3,3,2,2 \right]$}
\label{Fig4_5}

\begin{tikzpicture}[scale=0.85, auto=center, transform shape, >=triangle 45]
\tikzstyle{every node}=[draw,shape=circle];
  \node (n1) at (180:2) {};
  \node (n2) at (120:2) {};
  \node (n3) at (60:2)  {};
  \node (n4) at (0:2)   {};
  \node (n5) at (300:2) {};
  \node (n6) at (240:2) {};

  \foreach \from/\to in {n1/n3,n1/n4,n1/n5,n1/n6,n2/n5,n2/n6,n3/n5,n3/n6,n4/n6}
    \draw[dotted] (\from) -- (\to);
  \foreach \from/\to in {n1/n2,n2/n3,n2/n4,n3/n4,n4/n5,n5/n6}
    \draw[red,thick] (\from) -- (\to);
\end{tikzpicture}
\end{subfigure}
\begin{subfigure}{0.24\textwidth}
\centering
\caption{$G_6: \left[ 5,3,3,3,3,1 \right]$}
\label{Fig4_6}

\begin{tikzpicture}[scale=0.85, auto=center, transform shape, >=triangle 45]
\tikzstyle{every node}=[draw,shape=circle];
  \node (n1) at (180:2) {};
  \node (n2) at (120:2) {};
  \node (n3) at (60:2)  {};
  \node (n4) at (0:2)   {};
  \node (n5) at (300:2) {};
  \node (n6) at (240:2) {};

  \foreach \from/\to in {n1/n2,n1/n3,n1/n4,n1/n5,n1/n6,n3/n5,n3/n6,n4/n5,n4/n6,n5/n6}
    \draw[dotted] (\from) -- (\to);
  \foreach \from/\to in {n2/n3,n2/n4,n2/n5,n2/n6,n3/n4,n5/n6}
    \draw[red,thick] (\from) -- (\to);
\end{tikzpicture}
\end{subfigure}
\begin{subfigure}{0.24\textwidth}
\centering
\caption{$G_7: \left[ 5,3,3,3,2,2 \right]$}
\label{Fig4_7}

\begin{tikzpicture}[scale=0.85, auto=center, transform shape, >=triangle 45]
\tikzstyle{every node}=[draw,shape=circle];
  \node (n1) at (180:2) {};
  \node (n2) at (120:2) {};
  \node (n3) at (60:2)  {};
  \node (n4) at (0:2)   {};
  \node (n5) at (300:2) {};
  \node (n6) at (240:2) {};

  \foreach \from/\to in {n1/n2,n1/n3,n1/n4,n1/n5,n1/n6,n2/n4,n3/n5,n3/n6,n4/n6,n5/n6}
    \draw[dotted] (\from) -- (\to);
  \foreach \from/\to in {n2/n3,n2/n5,n2/n6,n3/n4,n4/n5,n5/n6}
    \draw[red,thick] (\from) -- (\to);
\end{tikzpicture}
\end{subfigure}
\begin{subfigure}{0.24\textwidth}
\centering
\caption{$G_8: \left[ 4,3,3,3,3,2 \right]$}
\label{Fig4_8}

\begin{tikzpicture}[scale=0.85, auto=center, transform shape, >=triangle 45]
\tikzstyle{every node}=[draw,shape=circle];
  \node (n1) at (180:2) {};
  \node (n2) at (120:2) {};
  \node (n3) at (60:2)  {};
  \node (n4) at (0:2)   {};
  \node (n5) at (300:2) {};
  \node (n6) at (240:2) {};

  \foreach \from/\to in {n1/n3,n1/n4,n1/n5,n1/n6,n2/n4,n2/n5,n3/n5,n3/n6,n4/n6,n5/n6}
    \draw[dotted] (\from) -- (\to);
  \foreach \from/\to in {n1/n2,n2/n3,n2/n6,n3/n4,n4/n5,n5/n6}
    \draw[red,thick] (\from) -- (\to);
\end{tikzpicture}
\end{subfigure}

\vspace{0.17cm}
\begin{subfigure}{0.24\textwidth}
\centering
\caption{$G_9: \left[ 5,4,3,3,2,1 \right]$}
\label{Fig4_9}

\begin{tikzpicture}[scale=0.85, auto=center, transform shape, >=triangle 45]
\tikzstyle{every node}=[draw,shape=circle];
  \node (n1) at (180:2) {};
  \node (n2) at (120:2) {};
  \node (n3) at (60:2)  {};
  \node (n4) at (0:2)   {};
  \node (n5) at (300:2) {};
  \node (n6) at (240:2) {};

  \foreach \from/\to in {n1/n2,n1/n3,n1/n4,n1/n5,n1/n6,n2/n4,n2/n6,n3/n6,n4/n6}
    \draw[dotted] (\from) -- (\to);
  \foreach \from/\to in {n2/n3,n2/n5,n3/n4,n3/n5,n4/n5,n5/n6}
    \draw[red,thick] (\from) -- (\to);
\end{tikzpicture}
\end{subfigure}
\begin{subfigure}{0.24\textwidth}
\centering
\caption{$G_{10}: \left[ 4,4,4,2,2,2 \right]$}
\label{Fig4_10}

\begin{tikzpicture}[scale=0.85, auto=center, transform shape, >=triangle 45]
\tikzstyle{every node}=[draw,shape=circle];
  \node (n1) at (180:2) {};
  \node (n2) at (120:2) {};
  \node (n3) at (60:2)  {};
  \node (n4) at (0:2)   {};
  \node (n5) at (300:2) {};
  \node (n6) at (240:2) {};

  \foreach \from/\to in {n1/n3,n1/n4,n1/n5,n1/n6,n2/n4,n2/n6,n3/n6,n4/n5,n4/n6}
    \draw[dotted] (\from) -- (\to);
  \foreach \from/\to in {n1/n2,n2/n3,n2/n5,n3/n4,n3/n5,n5/n6}
    \draw[red,thick] (\from) -- (\to);
\end{tikzpicture}
\end{subfigure}
\begin{subfigure}{0.24\textwidth}
\centering
\caption{$G_{11}: \left[ 4,4,4,3,2,1 \right]$}
\label{Fig4_11}

\begin{tikzpicture}[scale=0.85, auto=center, transform shape, >=triangle 45]
\tikzstyle{every node}=[draw,shape=circle];
  \node (n1) at (180:2) {};
  \node (n2) at (120:2) {};
  \node (n3) at (60:2)  {};
  \node (n4) at (0:2)   {};
  \node (n5) at (300:2) {};
  \node (n6) at (240:2) {};

  \foreach \from/\to in {n1/n3,n1/n4,n1/n5,n1/n6,n2/n4,n3/n6,n4/n5,n4/n6,n5/n6}
    \draw[dotted] (\from) -- (\to);
  \foreach \from/\to in {n1/n2,n2/n3,n2/n5,n2/n6,n3/n4,n3/n5}
    \draw[red,thick] (\from) -- (\to);
\end{tikzpicture}
\end{subfigure}
\begin{subfigure}{0.24\textwidth}
\centering
\caption{$G_{12}: \left[ 5,3,3,3,2,2 \right]$}
\label{Fig4_12}

\begin{tikzpicture}[scale=0.85, auto=center, transform shape, >=triangle 45]
\tikzstyle{every node}=[draw,shape=circle];
  \node (n1) at (180:2) {};
  \node (n2) at (120:2) {};
  \node (n3) at (60:2)  {};
  \node (n4) at (0:2)   {};
  \node (n5) at (300:2) {};
  \node (n6) at (240:2) {};

  \foreach \from/\to in {n1/n2,n1/n3,n1/n4,n1/n5,n1/n6,n2/n4,n3/n5,n3/n6,n5/n6}
    \draw[dotted] (\from) -- (\to);
  \foreach \from/\to in {n2/n3,n2/n5,n2/n6,n3/n4,n4/n5,n4/n6}
    \draw[red,thick] (\from) -- (\to);
\end{tikzpicture}
\end{subfigure}

\vspace{0.17cm}
\begin{subfigure}{0.24\textwidth}
\centering
\caption{$G_{13}: \left[ 4,4,3,3,3,1 \right]$}
\label{Fig4_13}

\begin{tikzpicture}[scale=0.85, auto=center, transform shape, >=triangle 45]
\tikzstyle{every node}=[draw,shape=circle];
  \node (n1) at (180:2) {};
  \node (n2) at (120:2) {};
  \node (n3) at (60:2)  {};
  \node (n4) at (0:2)   {};
  \node (n5) at (300:2) {};
  \node (n6) at (240:2) {};

  \foreach \from/\to in {n1/n3,n1/n4,n1/n5,n1/n6,n2/n4,n3/n5,n3/n6,n4/n6,n5/n6}
    \draw[dotted] (\from) -- (\to);
  \foreach \from/\to in {n1/n2,n2/n3,n2/n5,n2/n6,n3/n4,n4/n5}
    \draw[red,thick] (\from) -- (\to);
\end{tikzpicture}
\end{subfigure}
\begin{subfigure}{0.24\textwidth}
\centering
\caption{$G_{14}: \left[ 4,4,3,3,3,1 \right]$}
\label{Fig4_14}

\begin{tikzpicture}[scale=0.85, auto=center, transform shape, >=triangle 45]
\tikzstyle{every node}=[draw,shape=circle];
  \node (n1) at (180:2) {};
  \node (n2) at (120:2) {};
  \node (n3) at (60:2)  {};
  \node (n4) at (0:2)   {};
  \node (n5) at (300:2) {};
  \node (n6) at (240:2) {};

  \foreach \from/\to in {n1/n3,n1/n4,n1/n5,n1/n6,n2/n6,n3/n5,n3/n6,n4/n5,n4/n6}
    \draw[dotted] (\from) -- (\to);
  \foreach \from/\to in {n1/n2,n2/n3,n2/n4,n2/n5,n3/n4,n5/n6}
    \draw[red,thick] (\from) -- (\to);
\end{tikzpicture}
\end{subfigure}
\begin{subfigure}{0.24\textwidth}
\centering
\caption{$G_{15}: \left[ 4,4,3,3,2,2 \right]$}
\label{Fig4_15}

\begin{tikzpicture}[scale=0.85, auto=center, transform shape, >=triangle 45]
\tikzstyle{every node}=[draw,shape=circle];
  \node (n1) at (180:2) {};
  \node (n2) at (120:2) {};
  \node (n3) at (60:2)  {};
  \node (n4) at (0:2)   {};
  \node (n5) at (300:2) {};
  \node (n6) at (240:2) {};

  \foreach \from/\to in {n1/n2,n1/n3,n1/n4,n1/n5,n2/n4,n2/n6,n3/n6,n4/n6,n5/n6}
    \draw[dotted] (\from) -- (\to);
  \foreach \from/\to in {n1/n6,n2/n3,n2/n5,n3/n4,n3/n5,n4/n5}
    \draw[red,thick] (\from) -- (\to);
\end{tikzpicture}
\end{subfigure}
\begin{subfigure}{0.24\textwidth}
\centering
\caption{$G_{16}: \left[ 4,3,3,3,3,2 \right]$}
\label{Fig4_16}

\begin{tikzpicture}[scale=0.85, auto=center, transform shape, >=triangle 45]
\tikzstyle{every node}=[draw,shape=circle];
  \node (n1) at (180:2) {};
  \node (n2) at (120:2) {};
  \node (n3) at (60:2)  {};
  \node (n4) at (0:2)   {};
  \node (n5) at (300:2) {};
  \node (n6) at (240:2) {};

  \foreach \from/\to in {n1/n2,n1/n3,n1/n4,n1/n5,n2/n5,n2/n6,n3/n5,n3/n6,n4/n6}
    \draw[dotted] (\from) -- (\to);
  \foreach \from/\to in {n1/n6,n2/n3,n2/n4,n3/n4,n4/n5,n5/n6}
    \draw[red,thick] (\from) -- (\to);
\end{tikzpicture}
\end{subfigure}

\vspace{0.17cm}
\begin{subfigure}{0.24\textwidth}
\centering
\caption{$G_{17}: \left[ 4,4,3,3,2,2 \right]$}
\label{Fig4_17}

\begin{tikzpicture}[scale=0.85, auto=center, transform shape, >=triangle 45]
\tikzstyle{every node}=[draw,shape=circle];
  \node (n1) at (180:2) {};
  \node (n2) at (120:2) {};
  \node (n3) at (60:2)  {};
  \node (n4) at (0:2)   {};
  \node (n5) at (300:2) {};
  \node (n6) at (240:2) {};

  \foreach \from/\to in {n1/n2,n1/n3,n1/n4,n1/n6,n2/n5,n2/n6,n3/n5,n3/n6,n4/n6}
    \draw[dotted] (\from) -- (\to);
  \foreach \from/\to in {n1/n5,n2/n3,n2/n4,n3/n4,n4/n5,n5/n6}
    \draw[red,thick] (\from) -- (\to);
\end{tikzpicture}
\end{subfigure}
\begin{subfigure}{0.24\textwidth}
\centering
\caption{$G_{18}: \left[ 3,3,3,3,3,3 \right]$}
\label{Fig4_18}

\begin{tikzpicture}[scale=0.85, auto=center, transform shape, >=triangle 45]
\tikzstyle{every node}=[draw,shape=circle];
  \node (n1) at (180:2) {};
  \node (n2) at (120:2) {};
  \node (n3) at (60:2)  {};
  \node (n4) at (0:2)   {};
  \node (n5) at (300:2) {};
  \node (n6) at (240:2) {};

  \foreach \from/\to in {n1/n3,n1/n4,n1/n5,n2/n4,n2/n5,n2/n6,n3/n5,n3/n6,n4/n6}
    \draw[dotted] (\from) -- (\to);
  \foreach \from/\to in {n1/n2,n1/n6,n2/n3,n3/n4,n4/n5,n5/n6}
    \draw[red,thick] (\from) -- (\to);
\end{tikzpicture}
\end{subfigure}
\begin{subfigure}{0.24\textwidth}
\centering
\caption{$G_{19}: \left[ 5,5,2,2,2,2 \right]$}
\label{Fig4_19}

\begin{tikzpicture}[scale=0.85, auto=center, transform shape, >=triangle 45]
\tikzstyle{every node}=[draw,shape=circle];
  \node (n1) at (180:2) {};
  \node (n2) at (120:2) {};
  \node (n3) at (60:2)  {};
  \node (n4) at (0:2)   {};
  \node (n5) at (300:2) {};
  \node (n6) at (240:2) {};

  \foreach \from/\to in {n1/n2,n1/n3,n1/n4,n1/n5,n1/n6,n2/n6,n3/n6,n4/n6,n5/n6}
    \draw[dotted] (\from) -- (\to);
  \foreach \from/\to in {n2/n3,n2/n4,n2/n5,n3/n4,n3/n5,n4/n5}
    \draw[red,thick] (\from) -- (\to);
\end{tikzpicture}
\end{subfigure}
\begin{subfigure}{0.24\textwidth}
\centering
\caption{$G_{20}: \left[ 3,3,3,3,3,3 \right]$}
\label{Fig4_20}

\begin{tikzpicture}[scale=0.85, auto=center, transform shape, >=triangle 45]
\tikzstyle{every node}=[draw,shape=circle];
  \node (n1) at (180:2) {};
  \node (n2) at (120:2) {};
  \node (n3) at (60:2)  {};
  \node (n4) at (0:2)   {};
  \node (n5) at (300:2) {};
  \node (n6) at (240:2) {};

  \foreach \from/\to in {n1/n2,n1/n3,n1/n4,n2/n5,n2/n6,n3/n5,n3/n6,n4/n5,n4/n6}
    \draw[dotted] (\from) -- (\to);
  \foreach \from/\to in {n1/n5,n1/n6,n2/n3,n2/n4,n3/n4,n5/n6}
    \draw[red,thick] (\from) -- (\to);
\end{tikzpicture}
\end{subfigure}

\captionsetup{justification=centerfirst}
\caption{Possible graph representations of incomplete pairwise comparison matrices of size six with six missing entries  \\ \vspace{0.2cm}
\footnotesize{\emph{Notes}: Dotted lines indicate the known comparisons, thick red lines indicate the unknown comparisons. \\
The vectors show the degree distribution of the associated graph.
See Table~\ref{Table_A2} for the spectral radii and random indices of these incomplete pairwise comparison matrices.}}
\label{Fig4}
\end{figure}
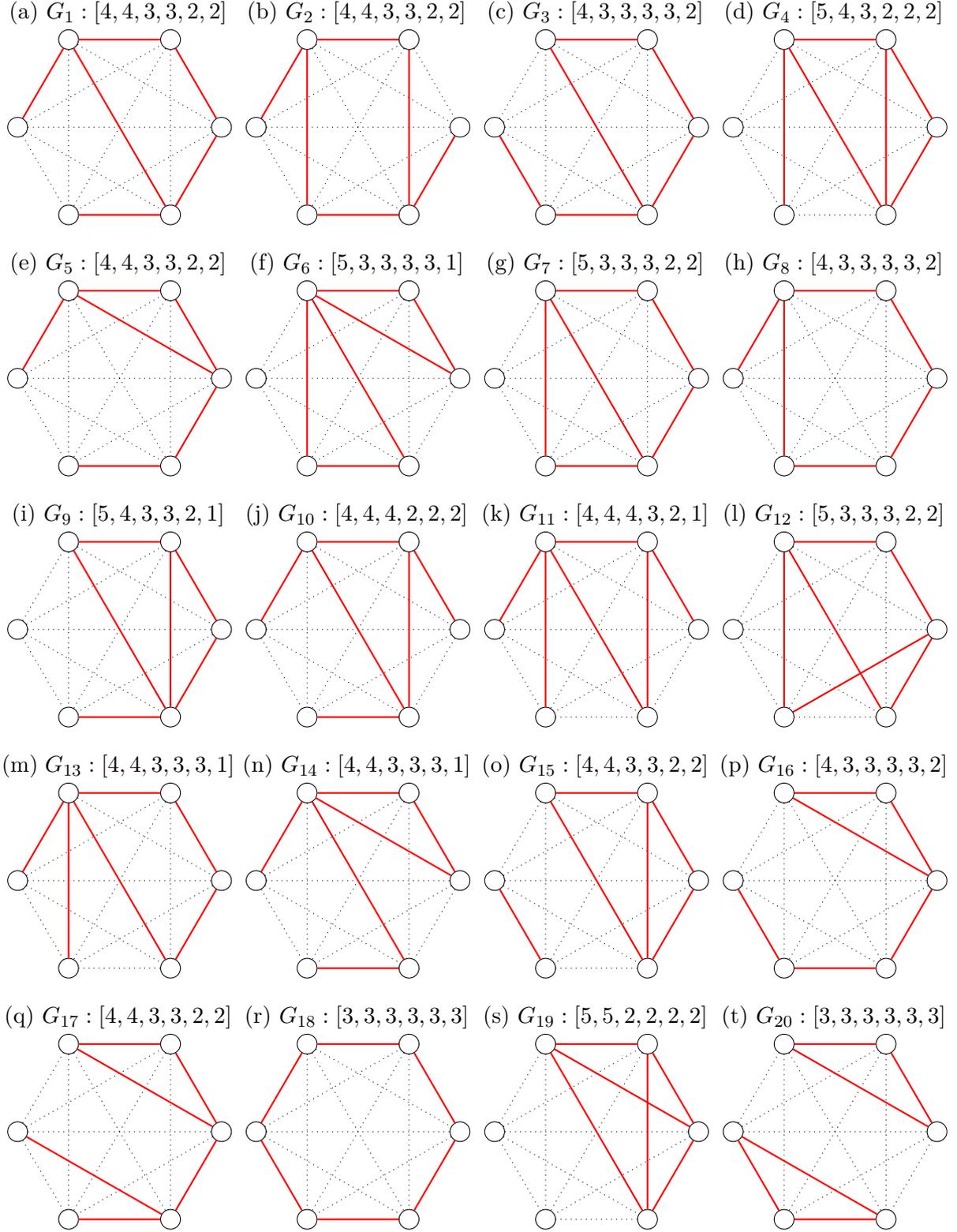


If $n=6$, we have computed the random indices only up to $m=7$ as the number of different graphs increases with $m$; for instance, 20 scenarios exist for $n=6$ and $m=6$ according to Figure~\ref{Fig4}.
The random indices are provided in Table~\ref{Table_A2} in the Appendix. Again, the regular graph $G_{20}$ in Figure~\ref{Fig4} occurs with the smallest probability and has the smallest value of $\mathit{RI}$ if $m=6$, and the other regular graph $G_{18}$ has the second smallest random index.

\begin{figure}[t!]
\centering

\begin{subfigure}{\textwidth}
\caption{$n=5$ alternatives}
\label{Fig5a}

\begin{tikzpicture}
\begin{axis}[
xlabel = Spectral radius,
x label style = {font=\small},
ylabel = Random index $\mathit {RI}$,
y label style = {font=\small},
width = \textwidth,
height = 0.6\textwidth,
ymajorgrids = true,
ymin = 0,
legend style = {font=\small,at={(0.15,-0.15)},anchor=north west,legend columns=6},
legend entries = {$m = 1 \qquad$,$m = 2 \qquad$,$m = 3 \qquad$,$m = 4 \qquad$,$m = 5$}
] 
\addplot [green, thick, only marks, mark=pentagon, mark size=2pt, mark options=solid] coordinates {
(3.645751,0.9246087)
};
\addplot [orange, thick, only marks, mark=asterisk, mark size=2pt, mark options=solid] coordinates {
(3.3234043,0.7453902)
(3.236068,0.7275285)
};
\addplot [blue, thick, only marks, mark=o, mark size=2pt, mark options={solid,semithick}] coordinates {
(3.0861302,0.5926127)
(2.9354323,0.5534586)
(2.855773,0.5376674)
(3,0.5610572)
};
\addplot [black, thick, only marks, mark=square, mark size=2pt, mark options={solid,thin}] coordinates {
(2.44949,0.342611)
(2.481194,0.3557099)
(2.561553,0.3744518)
(2.685544,0.3927302)
(2.6411865,0.3951718)
};
\addplot [ForestGreen, thick, only marks, mark=star, mark size=2pt, mark options=solid] coordinates {
(2.342923,0.2189586)
(2.302776,0.2235202)
(2.1357792,0.1899221)
(2.2143197,0.225597)
(2,0.1717379)
};
\end{axis}
\end{tikzpicture}
\end{subfigure}

\vspace{0.25cm}
\begin{subfigure}{\textwidth}
\caption{$n=6$ alternatives}
\label{Fig5b}

\begin{tikzpicture}
\begin{axis}[
xlabel = Spectral radius,
x label style = {font=\small},
ylabel = Random index $\mathit {RI}$,
y label style = {font=\small},
width = \textwidth,
height = 0.6\textwidth,
ymajorgrids = true,
ymin = 0,
legend style = {font=\small,at={(0,-0.15)},anchor=north west,legend columns=7},
legend entries = {$m = 1 \qquad$,$m = 2 \qquad$,$m = 3 \qquad$,$m = 4 \qquad$,$m = 5 \qquad$,$m = 6 \qquad$,$m = 7$}
] 
\addplot [green, thick, only marks, mark=pentagon, mark size=2pt, mark options=solid] coordinates {
(4.701562,1.127994)
};
\addplot [orange, thick, only marks, mark=asterisk, mark size=2pt, mark options=solid] coordinates {
(4.4278789,1.009885)
(4.372281,1.000693)
};
\addplot [blue, thick, only marks, mark=o, mark size=2pt, mark options={solid,semithick}] coordinates {
(4.2014723,0.8983255)
(4.1190267,0.8841247)
(4,0.8658453)
(4.162278,0.8904032)
(4.067786,0.8765042)
};
\addplot [black, thick, only marks, mark=square, mark size=2pt, mark options={solid,thin}] coordinates {
(4.051374,0.8059335)
(3.7320508,0.7457258)
(3.766435,0.749842)
(3.8589514,0.7658996)
(3.778457,0.7524769)
(3.713585,0.7430935)
(3.8200894,0.7597318)
(3.895107,0.7700051)
(3.828427,0.7600201)
};
\addplot [ForestGreen, thick, only marks, mark=star, mark size=2pt, mark options=solid] coordinates {
(3.467932,0.6267476)
(3.5344185,0.6356367)
(3.5926147,0.6428692)
(3.4978536,0.6296739)
(3.710452,0.6694259)
(3.514061,0.6310027)
(3.44949,0.6219347)
(3.3884897,0.6140633)
(3.626198,0.6464161)
(3.460867,0.6235273)
(3.6903064,0.6702097)
(3.372281,0.6125953)
(3.5615528,0.6412401)
(3.392344,0.6126941)
};
\addplot [brown, thick, only marks, mark=triangle, mark size=2pt, mark options=solid] coordinates {
(3.1819433,0.5055185)
(3.236068,0.5111271)
(3.0867986,0.4914545)
(3.281394,0.5130386)
(3.1691946,0.4993666)
(3.323404,0.5258631)
(3.2227433,0.5069287)
(3.1149075,0.4933358)
(3.4036819,0.5388832)
(3.236068,0.5097174)
(3.383931,0.5392852)
(3.261802,0.5168979)
(3.353856,0.5401792)
(3.294791,0.526741)
(3.141336,0.4940096)
(3.092166,0.4881983)
(3.188757,0.5022569)
(3,0.4782366)
(3.372281,0.5221835)
(3,0.4732735)
};
\addplot [red, thick, only marks, mark=diamond, mark size=2pt, mark options=solid] coordinates {
(2.980937,0.4029916)
(3.0143257,0.4015935)
(3.1642479,0.4293091)
(2.895107,0.3919521)
(2.943879,0.3909285)
(2.8529174,0.3759963)
(2.791288,0.3706638)
(2.8136065,0.3705947)
(3.102009,0.4055886)
(2.9326678,0.3903223)
(2.903212,0.3782781)
(3.0965072,0.4312295)
(3.047799,0.4081606)
(3.0437379,0.4003218)
(2.8422357,0.3846909)
(2.796437,0.3647622)
(2.732051,0.3681584)
(3.17741,0.4272843)
(2.741082,0.3598008)
(2.732051,0.3561878)
(2.828427,0.3633108)
(2.947413,0.3864188)
};
\end{axis}
\end{tikzpicture}
\end{subfigure}

\captionsetup{justification=centering}
\caption{Spectral radius and random index for incomplete matrices of size five and six}
\label{Fig5}

\end{figure}
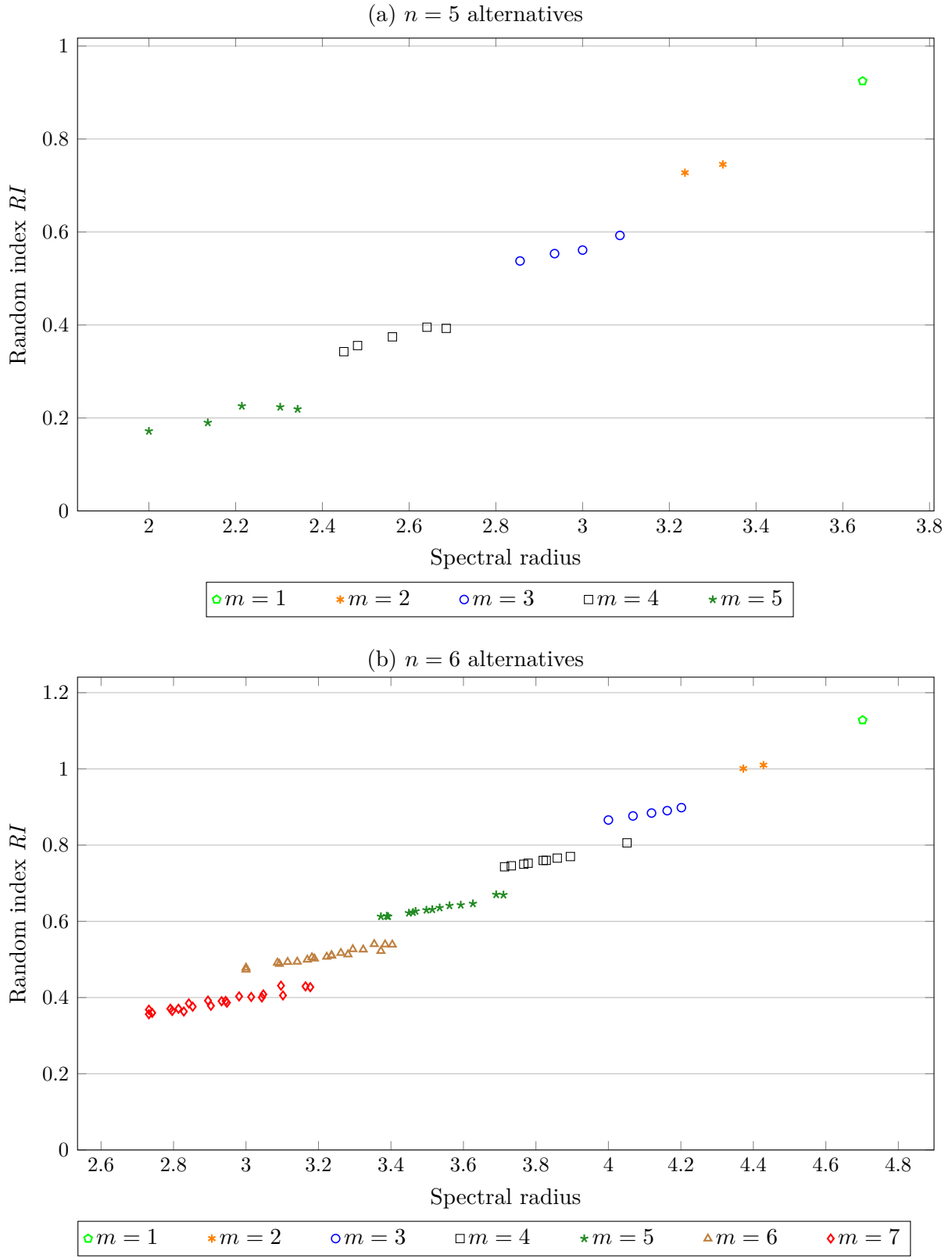


Figure~\ref{Fig5} presents the random indices for incomplete pairwise comparison matrices of size five (Figure~\ref{Fig5a}) and six (Figure~\ref{Fig5b}), as a function of the spectral radius of the associated graph. The monotonic relationship between the spectral radius and $\mathit{RI}$ breaks first when $n=5$ and $m=4$ (see also Table~\ref{Table_A1}). However, a higher spectral radius usually implies a higher random index even for $n=5$ and $n=6$. If the number of alternatives is six, the spectral radii can overlap for different values of missing entries $m$, but the random indices still monotonically decrease as a function of $m$. In other words, the number of variables in the optimisation problem~\eqref{EV_min} is more important than their relative position in the representing graph with respect to the optimum, which is intuitive.

\begin{table}[t!]
  \centering
  \caption{Extreme random indices for incomplete pairwise comparison \\ matrices of a fixed size $n$ and number of missing entries $m$}
  \label{Table2}
\rowcolors{3}{}{gray!20}
    \begin{tabularx}{0.6\textwidth}{ccCCC} \toprule
    $n$     & $m$     & Minimal $\mathit{RI}$ & Maximal $\mathit{RI}$ & Ratio \\ \bottomrule
    4     & 2     & 0.2646 & 0.3165 & 83.61\% \\ \hline
    5     & 2     & 0.7275 & 0.7454 & 97.60\% \\
    5     & 3     & 0.5377 & 0.5926 & 90.73\% \\
    5     & 4     & 0.3426 & 0.3952 & 86.70\% \\
    5     & 5     & 0.1717 & 0.2256 & 76.13\% \\ \hline
    6     & 2     & 1.0007 & 1.0099 & 99.09\% \\
    6     & 3     & 0.8658 & 0.8983 & 96.38\% \\
    6     & 4     & 0.7431 & 0.8059 & 92.20\% \\
    6     & 5     & 0.6126 & 0.6702 & 91.40\% \\
    6     & 6     & 0.4733 & 0.5402 & 87.61\% \\
    6     & 7     & 0.3562 & 0.4312 & 82.60\% \\ \toprule
    \end{tabularx}
\end{table}

Table~\ref{Table2} shows the extreme values of the random index $\mathit{RI}$ for given pairs of parameters $m$ and $n$. Unsurprisingly, the influence of the graph decreases if $m$ is fixed and the number of alternatives grows, but increases if $n$ is fixed and the number of missing entries increases. Among the 11 scenarios, the random index can be smaller by more than 10\% in five cases, and the reduction may exceed 23\% if both parameters are equal to five. This reinforces that the underlying graph cannot be ignored in the calculation of inconsistency thresholds for incomplete pairwise comparison matrices.

\begin{table}[t!]
  \centering
  \caption{Spectral radius and the distribution of triads for graphs representing \\ incomplete pairwise comparison matrices of size $n=6$ with $m=6$ missing entries}
  \label{Table3}
\rowcolors{3}{gray!20}{}
\begin{threeparttable}
    \begin{tabularx}{0.8\textwidth}{ccCCCC} \toprule \hiderowcolors
    \multirow{2}[0]{*}{Graph} & \multicolumn{1}{r}{\multirow{2}[0]{*}{Spectral radius}} & \multicolumn{4}{c}{Number of triads with missing entries $\mu$} \\
          &       & \multicolumn{1}{c}{$\mu = 0$} & \multicolumn{1}{c}{$\mu = 1$} & \multicolumn{1}{c}{$\mu = 2$} & \multicolumn{1}{c}{$\mu = 3$} \\ \bottomrule \showrowcolors
    $G_{1}$ & 3.1819 & 4     & 8     & 8     & 0 \\
    $G_{2}$ & 3.2361 & 4     & 8     & 8     & 0 \\
    $G_{3}$ & 3.0868 & 3     & 10    & 7     & 0 \\
    $G_{4}$ & 3.2814 & 4     & 10    & 4     & 2 \\
    $G_{5}$ & 3.1692 & 3     & 11    & 5     & 1 \\
    $G_{6}$ & 3.3234 & 4     & 10    & 4     & 2 \\
    $G_{7}$ & 3.2227 & 4     & 9     & 6     & 1 \\
    $G_{8}$ & 3.1149 & 3     & 10    & 7     & 0 \\
    $G_{9}$ & 3.4037 & 5     & 8     & 5     & 2 \\
    $G_{10}$ & 3.2361 & 4     & 9     & 6     & 1 \\
    $G_{11}$ & 3.3839 & 5     & 7     & 7     & 1 \\
    $G_{12}$ & 3.2618 & 5     & 6     & 9     & 0 \\
    $G_{13}$ & 3.3539 & 5     & 6     & 9     & 0 \\
    $G_{14}$ & 3.2948 & 4     & 9     & 6     & 1 \\
    $G_{15}$ & 3.1413 & 2     & 14    & 2     & 2 \\
    $G_{16}$ & 3.0922 & 2     & 13    & 4     & 1 \\
    $G_{17}$ & 3.1888 & 3     & 11    & 5     & 1 \\
    $G_{18}$ & 3     & 2     & 12    & 6     & 0 \\
    $G_{19}$ & 3.3723 & 4     & 12    & 0     & 4 \\
    $G_{20}$ & 3     & 0     & 18    & 0     & 2 \\ \toprule
    \end{tabularx}
\begin{tablenotes} \footnotesize
	\item \emph{Notes}: Each incomplete pairwise comparison matrix has 20 triads.
	\item The graphs are depicted in Figure~\ref{Fig4}.
\end{tablenotes}
\end{threeparttable}
\end{table}

The strong association between the spectral radius of the representing graph and the random index seems to be unexpected. Hence, we provide some theoretical intuition for this relation.
Table~\ref{Table3} presents the number of different triads (pairwise comparison matrices of size three) for the 20 graphs with $n=6$ and $m=6$ that can be seen in Figure~\ref{Fig4}. Six alternatives imply 20 triads in all cases, but these triads are quite different with respect to the number of missing entries $\mu$. For example, in graph $G_1$, there are 4 triads without missing entries, and 8-8 triads with one and two missing entries, respectively. On the other hand, graph $G_{20}$ has 18 triads with one missing entry, and 2 triads with three missing entries.

Inconsistency is closely connected to triads since the definition of consistency involves three alternatives. Indeed, several inconsistency indices simply aggregate the inconsistencies of all triads \citep{Brunelli2018}. Even though the inconsistency index $\mathit{CI}$ is not such a triad-based index, \citet[Table~2c]{Cavallo2020} uncovers an extremely strong monotonic increasing correlation between $\mathit{CI}$ and the inconsistency index suggested by \citet{PelaezLamata2003}, which is the average determinant of all sub-matrices of order three, for $4 \leq n \leq 7$. Furthermore, there exists a unique inconsistency index (up to multiplication) if $n=3$ \citep{Csato2019c}; hence, $\mathit{CI}$ is functionally related to any reasonable inconsistency index in the case of three alternatives.

Thus, reducing $\mathit{CI}$ (the objective function of problem~\eqref{EV_min}) is essentially equivalent to minimising the inconsistency of the triads. However, the inconsistency of a triad cannot be decreased if it has no missing entries ($\mu = 0$), and this is difficult due to the appearance of the same pairwise comparison in $n-2$ different triads if it has one missing entry ($\mu = 1$). Consequently, the optimum of optimisation problem~\eqref{EV_min} is expected to be lower if there are few triads with $\mu = 1$ and, especially, $\mu = 0$ missing entries.

In order to check the connection between the spectral radius and the distribution of triads, we have estimated two linear regressions by ordinary least squares (OLS). The dependent variable is the spectral radius, while the independent variables are the constant and the number of triads with $\mu = 0$ and $\mu = 1$ in both cases. The two samples are provided by graphs with $m=6$ (see Table~\ref{Table3}) and $m=7$ when the number of alternatives is $n=6$. In both regressions, all coefficients remain significantly positive even at the 0.1\% level, and the value of adjusted $R^2$ exceeds 0.9. Hence, the spectral radius basically depends on the number of these two types of triads.

To summarise, the value of the spectral radius is probably related to the minimisation of the dominant eigenvalue in~\eqref{EV_min} through the distribution of triads with respect to the number of missing entries $\mu$. A higher spectral radius is associated with more triads that have at most one missing entry, when decreasing the average inconsistency of triads becomes more challenging. In addition, the optimal value of $\mathit{CI}$ is almost functionally related to the average inconsistency of triads, as shown by \citet{Cavallo2020}.

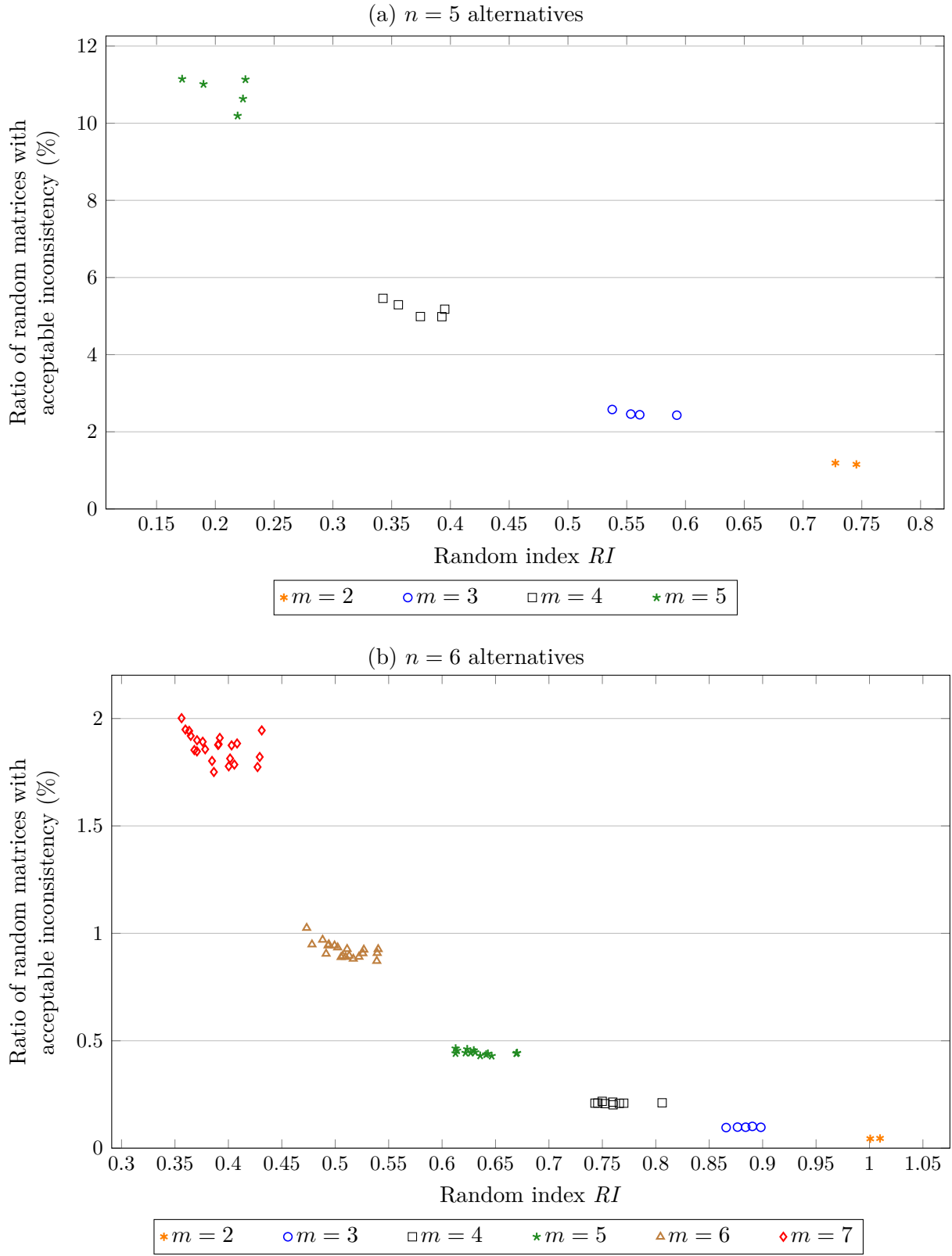
\begin{figure}[t!]
\centering

\begin{subfigure}{\textwidth}
\caption{$n=5$ alternatives}
\label{Fig6a}

\begin{tikzpicture}
\begin{axis}[
xlabel = Random index $\mathit {RI}$,
x label style = {font=\small},
ylabel = {Ratio of random matrices with \\ acceptable inconsistency (\%)},
y label style = {font=\small,align=center},
y tick label style = {/pgf/number format/.cd,fixed,precision=2},
width = 0.98\textwidth,
height = 0.6\textwidth,
ymajorgrids = true,
xmax = 0.82,
ymin = 0,
legend style = {font=\small,at={(0.2,-0.15)},anchor=north west,legend columns=6},
legend entries = {$m = 2 \qquad$,$m = 3 \qquad$,$m = 4 \qquad$,$m = 5$}
] 
\addplot [orange, thick, only marks, mark=asterisk, mark size=2pt, mark options=solid] coordinates {
(0.7453902,1.1555)
(0.7275285,1.1901)
};
\addplot [blue, thick, only marks, mark=o, mark size=2pt, mark options={solid,semithick}] coordinates {
(0.5926127,2.431)
(0.5534586,2.4598)
(0.5376674,2.5783)
(0.5610572,2.4421)
};
\addplot [black, thick, only marks, mark=square, mark size=2pt, mark options={solid,thin}] coordinates {
(0.342611,5.4596)
(0.3557099,5.2928)
(0.3744518,4.9879)
(0.3927302,4.9845)
(0.3951718,5.1779)
};
\addplot [ForestGreen, thick, only marks, mark=star, mark size=2pt, mark options=solid] coordinates {
(0.2189586,10.1918)
(0.2235202,10.6326)
(0.1899221,11.0126)
(0.225597,11.1338)
(0.1717379,11.146)
};
\end{axis}
\end{tikzpicture}
\end{subfigure}

\vspace{0.25cm}
\begin{subfigure}{\textwidth}
\caption{$n=6$ alternatives}
\label{Fig6b}

\begin{tikzpicture}
\begin{axis}[
xlabel = Random index $\mathit {RI}$,
x label style = {font=\small},
ylabel = {Ratio of random matrices with \\ acceptable inconsistency (\%)},
y label style = {font=\small,align=center},
y tick label style = {/pgf/number format/.cd,fixed,precision=2},
width = 0.98\textwidth,
height = 0.6\textwidth,
ymajorgrids = true,
ymin = 0,
legend style = {font=\small,at={(0.05,-0.15)},anchor=north west,legend columns=7},
legend entries = {$m = 2 \qquad$,$m = 3 \qquad$,$m = 4 \qquad$,$m = 5 \qquad$,$m = 6 \qquad$,$m = 7$}
] 
\addplot [orange, thick, only marks, mark=asterisk, mark size=2pt, mark options=solid] coordinates {
(1.009885,0.0454)
(1.000693,0.0439)
};
\addplot [blue, thick, only marks, mark=o, mark size=2pt, mark options={solid,semithick}] coordinates {
(0.8983255,0.0972)
(0.8841247,0.0975)
(0.8658453,0.0959)
(0.8904032,0.1014)
(0.8765042,0.098)
};
\addplot [black, thick, only marks, mark=square, mark size=2pt, mark options={solid,thin}] coordinates {
(0.8059335,0.2112)
(0.7457258,0.21)
(0.749842,0.218)
(0.7658996,0.2086)
(0.7524769,0.2073)
(0.7430935,0.2096)
(0.7597318,0.2143)
(0.7700051,0.2095)
(0.7600201,0.2018)
};
\addplot [ForestGreen, thick, only marks, mark=star, mark size=2pt, mark options=solid] coordinates {
(0.6267476,0.4434)
(0.6356367,0.4306)
(0.6428692,0.4402)
(0.6296739,0.4553)
(0.6694259,0.4409)
(0.6310027,0.4465)
(0.6219347,0.4439)
(0.6140633,0.4542)
(0.6464161,0.4294)
(0.6235273,0.461)
(0.6702097,0.4435)
(0.6125953,0.4415)
(0.6412401,0.4362)
(0.6126941,0.4643)
};
\addplot [brown, thick, only marks, mark=triangle, mark size=2pt, mark options=solid] coordinates {
(0.5055185,0.8893)
(0.5111271,0.9269)
(0.4914545,0.9049)
(0.5130386,0.8959)
(0.4993666,0.9433)
(0.5258631,0.907)
(0.5069287,0.8945)
(0.4933358,0.9433)
(0.5388832,0.8714)
(0.5097174,0.89)
(0.5392852,0.9083)
(0.5168979,0.8816)
(0.5401792,0.9262)
(0.526741,0.9241)
(0.4940096,0.9486)
(0.4881983,0.9699)
(0.5022569,0.9341)
(0.4782366,0.9481)
(0.5221835,0.8906)
(0.4732735,1.0252)
};
\addplot [red, thick, only marks, mark=diamond, mark size=2pt, mark options=solid] coordinates {
(0.4029916,1.8749)
(0.4015935,1.8138)
(0.4293091,1.8208)
(0.3919521,1.9096)
(0.3909285,1.8802)
(0.3759963,1.8917)
(0.3706638,1.8988)
(0.3705947,1.847)
(0.4055886,1.7856)
(0.3903223,1.8771)
(0.3782781,1.8568)
(0.4312295,1.9445)
(0.4081606,1.8838)
(0.4003218,1.7774)
(0.3846909,1.8023)
(0.3647622,1.9192)
(0.3681584,1.8531)
(0.4272843,1.774)
(0.3598008,1.9485)
(0.3561878,2.0014)
(0.3633108,1.9422)
(0.3864188,1.751)
};
\end{axis}
\end{tikzpicture}
\end{subfigure}

\captionsetup{justification=centering}
\caption{Random index and the probability of acceptable \\ inconsistency for incomplete matrices of size five and six}
\label{Fig6}

\end{figure}


Finally, Figure~\ref{Fig6} presents the relationship between the random index and the proportion of randomly generated incomplete pairwise comparison matrices with the same graph representation that have an acceptable inconsistency. The effect of the graph becomes more important if the number of missing entries $m$ is higher. Interestingly, a smaller (more restrictive) random index may lead to a higher probability of acceptable inconsistency, similar to what can be seen in Table~\ref{Table1} for $n=4$ alternatives.

\subsection{An approximation formula for the random index} \label{Sec34}

The computations in Section~\ref{Sec33} are quite complex and time-consuming, which calls for a heuristic predictor of the random value based on the spectral radius $\rho$.

Let $\mathcal{G}(n,m)$ be the set of graphs for a given number of alternatives $n$ and missing comparisons $m$. Let $\rho_{n,m}$ and $\mathit{RI}_{n,m}$ be the expected values of the spectral radius and random index for a fixed pair of $n$ and $m$, respectively.
For any graph $G_i \in \mathcal{G}(n,m)$, define
\[
\Delta \rho_i = \rho \left( G_i \right) - \rho_{n,m} \text{, and}
\]
\[
\Delta \mathit{RI}_i = \mathit{RI} \left( G_i \right) - \mathit{RI}_{n,m}.
\]
We estimate the linear regression $\Delta \mathit{RI}_i = \beta \Delta \rho_i$ by ordinary least squares (OLS) for the 73 observations provided by $n=6$ and $1 \leq m \leq 7$ in Table~\ref{Table_A2} in the Appendix. This gives $\beta = 0.16119$.
However, it is not guaranteed that $\beta = 0.16119$ remains valid if the number of alternatives is $n \geq 7$. For example, $\beta = 0.19219$ if the sample is the set of $5 \times 5$ matrices, namely, Table~\ref{Table_A1}. Indeed, the regression line is apparently steeper in Figure~\ref{Fig5a} than in Figure~\ref{Fig5b}, but the number of observations is only 17 in the first case.
Thus, $\beta = 0.16119$ is assumed to hold for any $n \geq 7$ in the following.

Then, for any graph $G_i \in \mathcal{G}(n,m)$, the random index $\mathit{RI}_i$ is approximated by:
\begin{equation} \label{RI_approx_rho}
\mathit{RI}_i \approx \mathit{RI}_{n,m} + \beta \Delta \rho_i.
\end{equation}
In formula~\eqref{RI_approx_rho}, $\Delta \rho_i$ depends on $\rho_i$ (the spectral radius of graph $G_i$) and $\rho_{n,m}$. The latter is estimated by generating 100 thousand random pairwise comparison matrices for each pair of parameters $n$ and $m$, and computing the average spectral radius if the corresponding graph is connected, which is not guaranteed for a high number of missing comparisons $m$. These values are presented in Table~\ref{Table_A3} in the Appendix for $6 \leq n \leq 9$. Note that this calculation does not require any specific algorithm or solving any optimisation problem.

On the other hand, if $n \geq 7$, the values of $\mathit{RI}_{n,m}$ are neither provided here, nor in \citet{AgostonCsato2022} for all $m$. However, \citet{AgostonCsato2022} present an approximation based on $\mathit{RI}_n$ that is available for all plausible number of alternatives $n$ \citep[Table~1]{AguaronMoreno-Jimenez2003}:
\begin{equation} \label{RI_approx_m}
\mathit{RI}_{n,m} \approx \left[ 1 - \frac{2m}{(n-1)(n-2)} \right] \mathit{RI}_n.
\end{equation}

We have checked formula~\eqref{RI_approx_rho} for a pairwise comparison matrix with $n=8$ alternatives, $m=5$ missing entries, and a spectral radius of $\rho_i = 6.097$.
The expected value of the spectral radius is $\rho_{8,5} = 5.862$ (see Table~\ref{Table_A3} in the Appendix), which leads to $\Delta \rho_i = 0.235$. The value of $\mathit{RI}_{n,m}$ is estimated from~\eqref{RI_approx_m} as $1.07$ since $\mathit{RI}_n = 1.404$. Hence, our heuristic predictor results in $\mathit{RI}_i = 1.07 + 0.16119 \cdot 0.235 = 1.1076$. The correct random index, given by the methodology described in Section~\ref{Sec23}, equals $1.117$. Consequently, the error of the approximation is within 1\% in this particular case.

\section{Conclusions} \label{Sec4}

Our paper provides refined random indices for incomplete pairwise comparison matrices in order to determine the threshold of acceptability for their inconsistency. The values of \citet{AgostonCsato2022} are made more accurate by considering not only the number of alternatives and missing entries, but also the graph representing the known comparisons. We show that the graph structure has a non-negligible effect and the random index is in a strict---albeit not perfect---relationship with the spectral radius of the corresponding graph. Furthermore, the underlying graph influences the ratio of randomly generated pairwise comparison matrices whose inconsistency can be accepted, which might question the validity of the 10\% rule of Saaty.

The research is far from closed.
First, the random indices are fully computed only up to five alternatives due to the complexity of the numerical calculations.
Second, even though the association with the spectral radius is surprisingly strong, other measures of graph ``robustness'' may provide an even better fit. 

\section*{Acknowledgements}
\addcontentsline{toc}{section}{Acknowledgements}
\noindent
We acknowledge the Digital Government Development and Project Management Ltd.\ for awarding us access to the Komondor HPC facility based in Hungary. \\
We are grateful to \emph{Zsombor Sz\'adoczki} for useful advice. \\
Two reviewers and three anonymous colleagues provided valuable comments and suggestions on earlier drafts. \\
The research was supported by the National Research, Development and Innovation Office under Grants Advanced 152220 and FK 145838, and the J\'anos Bolyai Research Scholarship of the Hungarian Academy of Sciences.

\bibliographystyle{apalike}
\bibliography{All_references}

\clearpage
\section*{Appendix}

\setcounter{table}{0}
\renewcommand{\thetable}{A.\arabic{table}}

\begin{table}[ht!]
  \centering
  \caption{Random indices for incomplete pairwise comparison matrices of size $n=5$}
  \label{Table_A1}
\rowcolors{3}{}{gray!20}
\begin{threeparttable}
    \begin{tabularx}{\textwidth}{cc Rccr} \toprule
    Value of $m$   & Graph & Probability & Spectral radius & Random index $\mathit{RI}$ &  Ratio of acc. \\ \bottomrule
    1     & 1     & 100.00\% & 3.6458 & 0.9246 & \multicolumn{1}{r}{---} \\ \hline
    2     & 1     & 67.05\% & 3.3234 & 0.7454 & 1.16\% \\
    2     & 2     & 32.95\% & 3.2361 & 0.7275 & 1.19\% \\ \hline
    3     & 1     & 16.92\% & 3.0861 & 0.5926 & 2.43\% \\
    3     & 2     & 49.49\% & 2.9354 & 0.5535 & 2.46\% \\
    3     & 3     & 25.28\% & 2.8558 & 0.5377 & 2.58\% \\
    3     & 4     & 8.31\% & 3     & 0.5611 & 2.44\% \\ \hline
    4     & 1     & 4.62\% & 2.4495 & 0.3426 & 5.46\% \\
    4     & 2     & 29.59\% & 2.4812 & 0.3557 & 5.29\% \\
    4     & 3     & 6.89\% & 2.5616 & 0.3745 & 4.99\% \\
    4     & 4     & 29.05\% & 2.6855 & 0.3927 & 4.98\% \\
    4     & 5     & 29.85\% & 2.6412 & 0.3952 & 5.18\% \\ \hline
    5     & 1     & 12.92\% & 2.3429 & 0.2190 & 10.19\% \\
    5     & 2     & 27.28\% & 2.3028 & 0.2235 & 10.63\% \\
    5     & 3     & 27.31\% & 2.1358 & 0.1899 & 11.01\% \\
    5     & 4     & 26.76\% & 2.2143 & 0.2256 & 11.13\% \\
    5     & 5     & 5.72\% & 2     & 0.1717 & 11.15\% \\ \toprule
    \end{tabularx}
\begin{tablenotes} \footnotesize
\item
\emph{Notes}: The column Probability shows the probability that the given graph occurs if all missing comparisons are placed randomly for a fixed pair of parameters $m$ and $n$.
\item
The column Ratio of acc.\ shows the probability that an incomplete pairwise comparison matrix with the given graph representation, whose known entries are drawn uniformly from the Saaty scale, has an acceptable level of inconsistency, for a fixed pair of parameters $m$ and $n$.
\item
If the spectral radius is equal to an integer $k$, the graph is $k$-regular.
\end{tablenotes}
\end{threeparttable}
\end{table}

\clearpage

\begin{ThreePartTable}
\begin{TableNotes}[normal] \footnotesize
\item
\emph{Notes}: The column Probability shows the probability that the given graph occurs if all missing comparisons are placed randomly for a fixed pair of parameters $m$ and $n$.
\item
The column Ratio of acc.\ shows the probability that an incomplete pairwise comparison matrix with the given graph representation, whose known entries are drawn uniformly from the Saaty scale, has an acceptable level of inconsistency, for a fixed pair of parameters $m$ and $n$.
\item
If the spectral radius is equal to an integer $k$, the graph is $k$-regular.
\end{TableNotes}

\rowcolors{2}{}{gray!20}
\begin{xltabular}{\textwidth}{cc Rccr}
\caption{Random indices for incomplete pairwise comparison matrices of size $n=6$}
\label{Table_A2} \\

\toprule
\multicolumn{1}{l}{Value of $m$} & \multicolumn{1}{c}{Graph} & \multicolumn{1}{R}{Probability} & \multicolumn{1}{c}{Spectral radius} & \multicolumn{1}{c}{Random index $\mathit{RI}$} & \multicolumn{1}{r}{Ratio of acc.} \\ \bottomrule
\endfirsthead

\multicolumn{5}{c}%
{\normalsize{\tablename\ \thetable{} (continued from the previous page)}} \vspace{0.2cm} \\ \toprule
\multicolumn{1}{l}{Value of $m$} & \multicolumn{1}{c}{Graph} & \multicolumn{1}{R}{Probability} & \multicolumn{1}{c}{Spectral radius} & \multicolumn{1}{c}{Random index $\mathit{RI}$} & \multicolumn{1}{r}{Ratio of acc.} \\ \bottomrule
\endhead

\toprule \hiderowcolors \multicolumn{6}{r}{{Continued on the next page}} \\ \showrowcolors \bottomrule
\endfoot

\toprule
\insertTableNotes
\endlastfoot

    1     & 1     & 100\% & 4.7016 & 1.1280 & \multicolumn{1}{r}{---} \\ \hline
    2     & 1     & 56.87\% & 4.4279 & 1.0099 & 0.05\% \\
    2     & 2     & 43.13\% & 4.3723 & 1.0007 & 0.04\% \\ \hline
    3     & 1     & 13.16\% & 4.2015 & 0.8983 & 0.10\% \\
    3     & 2     & 39.95\% & 4.1190 & 0.8841 & 0.10\% \\
    3     & 3     & 3.34\% & 4     & 0.8658 & 0.10\% \\
    3     & 4     & 4.39\% & 4.1623 & 0.8904 & 0.10\% \\
    3     & 5     & 39.16\% & 4.0678 & 0.8765 & 0.10\% \\ \hline
    4     & 1     & 2.10\% & 4.0514 & 0.8059 & 0.21\% \\
    4     & 2     & 6.60\% & 3.7321 & 0.7457 & 0.21\% \\
    4     & 3     & 4.50\% & 3.7664 & 0.7498 & 0.22\% \\
    4     & 4     & 26.75\% & 3.8590 & 0.7659 & 0.21\% \\
    4     & 5     & 25.76\% & 3.7785 & 0.7525 & 0.21\% \\
    4     & 6     & 13.58\% & 3.7136 & 0.7431 & 0.21\% \\
    4     & 7     & 4.31\% & 3.8201 & 0.7597 & 0.21\% \\
    4     & 8     & 13.12\% & 3.8951 & 0.7700 & 0.21\% \\
    4     & 9     & 3.28\% & 3.8284 & 0.7600 & 0.20\% \\ \hline
    5     & 1     & 12.48\% & 3.4679 & 0.6267 & 0.44\% \\
    5     & 2     & 12.35\% & 3.5344 & 0.6356 & 0.43\% \\
    5     & 3     & 11.74\% & 3.5926 & 0.6429 & 0.44\% \\
    5     & 4     & 12.15\% & 3.4979 & 0.6297 & 0.46\% \\
    5     & 5     & 6.00\% & 3.7105 & 0.6694 & 0.44\% \\
    5     & 6     & 11.81\% & 3.5141 & 0.6310 & 0.45\% \\
    5     & 7     & 2.07\% & 3.4495 & 0.6219 & 0.44\% \\
    5     & 8     & 11.80\% & 3.3885 & 0.6141 & 0.45\% \\
    5     & 9     & 3.17\% & 3.6262 & 0.6464 & 0.43\% \\
    5     & 10    & 6.05\% & 3.4609 & 0.6235 & 0.46\% \\
    5     & 11    & 3.92\% & 3.6903 & 0.6702 & 0.44\% \\
    5     & 12    & 1.51\% & 3.3723 & 0.6126 & 0.44\% \\
    5     & 13    & 3.00\% & 3.5616 & 0.6412 & 0.44\% \\
    5     & 14    & 1.95\% & 3.3923 & 0.6127 & 0.46\% \\ \hline
    6     & 1     & 7.38\% & 3.1819 & 0.5055 & 0.89\% \\
    6     & 2     & 3.92\% & 3.2361 & 0.5111 & 0.93\% \\
    6     & 3     & 7.48\% & 3.0868 & 0.4915 & 0.90\% \\
    6     & 4     & 7.24\% & 3.2814 & 0.5130 & 0.90\% \\
    6     & 5     & 14.59\% & 3.1692 & 0.4994 & 0.94\% \\
    6     & 6     & 2.01\% & 3.3234 & 0.5259 & 0.91\% \\
    6     & 7     & 6.87\% & 3.2227 & 0.5069 & 0.89\% \\
    6     & 8     & 6.92\% & 3.1149 & 0.4933 & 0.94\% \\
    6     & 9     & 7.18\% & 3.4037 & 0.5389 & 0.87\% \\
    6     & 10    & 2.36\% & 3.2361 & 0.5097 & 0.89\% \\
    6     & 11    & 7.14\% & 3.3839 & 0.5393 & 0.91\% \\
    6     & 12    & 1.44\% & 3.2618 & 0.5169 & 0.88\% \\
    6     & 13    & 3.28\% & 3.3539 & 0.5402 & 0.93\% \\
    6     & 14    & 7.54\% & 3.2948 & 0.5267 & 0.92\% \\
    6     & 15    & 1.93\% & 3.1413 & 0.4940 & 0.95\% \\
    6     & 16    & 7.34\% & 3.0922 & 0.4882 & 0.97\% \\
    6     & 17    & 3.64\% & 3.1888 & 0.5023 & 0.93\% \\
    6     & 18    & 1.23\% & 3     & 0.4782 & 0.95\% \\
    6     & 19    & 0.32\% & 3.3723 & 0.5222 & 0.89\% \\
    6     & 20    & 0.16\% & 3     & 0.4733 & 1.03\% \\ \hline
    7     & 1     & 10.93\% & 2.9809 & 0.4030 & 1.87\% \\
    7     & 2     & 11.70\% & 3.0143 & 0.4016 & 1.81\% \\
    7     & 3     & 2.87\% & 3.1642 & 0.4293 & 1.82\% \\
    7     & 4     & 2.92\% & 2.8951 & 0.3920 & 1.91\% \\
    7     & 5     & 6.32\% & 2.9439 & 0.3909 & 1.88\% \\
    7     & 6     & 11.43\% & 2.8529 & 0.3760 & 1.89\% \\
    7     & 7     & 2.95\% & 2.7913 & 0.3707 & 1.90\% \\
    7     & 8     & 5.84\% & 2.8136 & 0.3706 & 1.85\% \\
    7     & 9     & 2.10\% & 3.1020 & 0.4056 & 1.79\% \\
    7     & 10    & 5.95\% & 2.9327 & 0.3903 & 1.88\% \\
    7     & 11    & 3.07\% & 2.9032 & 0.3783 & 1.86\% \\
    7     & 12    & 1.88\% & 3.0965 & 0.4312 & 1.94\% \\
    7     & 13    & 2.87\% & 3.0478 & 0.4082 & 1.88\% \\
    7     & 14    & 5.59\% & 3.0437 & 0.4003 & 1.78\% \\
    7     & 15    & 2.83\% & 2.8422 & 0.3847 & 1.80\% \\
    7     & 16    & 6.34\% & 2.7964 & 0.3648 & 1.92\% \\
    7     & 17    & 2.82\% & 2.7321 & 0.3682 & 1.85\% \\
    7     & 18    & 0.87\% & 3.1774 & 0.4273 & 1.77\% \\
    7     & 19    & 5.76\% & 2.7411 & 0.3598 & 1.95\% \\
    7     & 20    & 1.62\% & 2.7321 & 0.3562 & 2.00\% \\
    7     & 21    & 0.28\% & 2.8284 & 0.3633 & 1.94\% \\
    7     & 22    & 3.07\% & 2.9474 & 0.3864 & 1.75\% \\
\end{xltabular}
\end{ThreePartTable}

\clearpage

\begin{ThreePartTable}
\begin{TableNotes}[normal] \footnotesize
\item
\emph{Notes}: 100 thousand random pairwise comparison matrices are generated for each pair of parameters $m$ and $n$, but the average spectral radius is computed only from connected graphs.
\item
The column Connected graphs shows the number of connected graphs for a fixed pair of parameters $m$ and $n$.
\end{TableNotes}

\rowcolors{2}{}{gray!20}
\begin{xltabular}{\textwidth}{cc CC}
\caption{Expected value of the spectral radius for graphs representing \\ incomplete pairwise comparison matrices of size $6 \leq n \leq 9$}
\label{Table_A3} \\

\toprule
Alternatives ($n$) & Missing comparisons ($m$) & Average spectral radius ($\rho_{n,m}$) & Connected graphs \\ \bottomrule
\endfirsthead

\multicolumn{4}{c}%
{\normalsize{\tablename\ \thetable{} (continued from the previous page)}} \vspace{0.2cm} \\ \toprule
Alternatives ($n$) & Missing comparisons ($m$) & Average spectral radius ($\rho_{n,m}$) & Connected graphs \\ \bottomrule
\endhead

\toprule \hiderowcolors \multicolumn{4}{r}{{Continued on the next page}} \\ \showrowcolors \bottomrule
\endfoot

\toprule
\insertTableNotes
\endlastfoot

    6     & 1     & 4.7016 & 100000 \\
    6     & 2     & 4.4040 & 100000 \\
    6     & 3     & 4.1074 & 100000 \\
    6     & 4     & 3.8124 & 100000 \\
    6     & 5     & 3.5177 & 99792 \\
    6     & 6     & 3.2217 & 98801 \\
    6     & 7     & 2.9213 & 95783 \\
    6     & 8     & 2.6127 & 88763 \\
    6     & 9     & 2.2896 & 75105 \\ \hline
    7     & 1     & 5.7417 & 100000 \\
    7     & 2     & 5.4837 & 100000 \\
    7     & 3     & 5.2263 & 100000 \\
    7     & 4     & 4.9694 & 100000 \\
    7     & 5     & 4.7129 & 100000 \\
    7     & 6     & 4.4571 & 99990 \\
    7     & 7     & 4.2014 & 99896 \\
    7     & 8     & 3.9463 & 99608 \\
    7     & 9     & 3.6899 & 98838 \\
    7     & 10    & 3.4315 & 97164 \\
    7     & 11    & 3.1712 & 93953 \\
    7     & 12    & 2.9050 & 88140 \\
    7     & 13    & 2.6308 & 78316 \\
    7     & 14    & 2.3457 & 63384 \\ \hline
    8     & 1     & 6.7720 & 100000 \\
    8     & 2     & 6.5442 & 100000 \\
    8     & 3     & 6.3166 & 100000 \\
    8     & 4     & 6.0891 & 100000 \\
    8     & 5     & 5.8621 & 100000 \\
    8     & 6     & 5.6351 & 100000 \\
    8     & 7     & 5.4085 & 100000 \\
    8     & 8     & 5.1821 & 99998 \\
    8     & 9     & 4.9564 & 99976 \\
    8     & 10    & 4.7312 & 99916 \\
    8     & 11    & 4.5057 & 99776 \\
    8     & 12    & 4.2802 & 99460 \\
    8     & 13    & 4.0543 & 98783 \\
    8     & 14    & 3.8271 & 97623 \\
    8     & 15    & 3.5985 & 95556 \\
    8     & 16    & 3.3669 & 92188 \\
    8     & 17    & 3.1330 & 87003 \\
    8     & 18    & 2.8932 & 79219 \\
    8     & 19    & 2.6464 & 67968 \\
    8     & 20    & 2.3937 & 53336 \\ \hline
    9     & 1     & 7.7958 & 100000 \\
    9     & 2     & 7.5918 & 100000 \\
    9     & 3     & 7.3879 & 100000 \\
    9     & 4     & 7.1841 & 100000 \\
    9     & 5     & 6.9803 & 100000 \\
    9     & 6     & 6.7768 & 100000 \\
    9     & 7     & 6.5732 & 100000 \\
    9     & 8     & 6.3699 & 100000 \\
    9     & 9     & 6.1669 & 100000 \\
    9     & 10    & 5.9641 & 99999 \\
    9     & 11    & 5.7615 & 99995 \\
    9     & 12    & 5.5589 & 99991 \\
    9     & 13    & 5.3568 & 99970 \\
    9     & 14    & 5.1547 & 99919 \\
    9     & 15    & 4.9525 & 99814 \\
    9     & 16    & 4.7507 & 99639 \\
    9     & 17    & 4.5486 & 99272 \\
    9     & 18    & 4.3461 & 98727 \\
    9     & 19    & 4.1433 & 97796 \\
    9     & 20    & 3.9390 & 96356 \\
    9     & 21    & 3.7332 & 94115 \\
    9     & 22    & 3.5260 & 90745 \\
    9     & 23    & 3.3160 & 85917 \\
    9     & 24    & 3.1026 & 79167 \\
    9     & 25    & 2.8848 & 69962 \\
    9     & 26    & 2.6613 & 58357 \\
    9     & 27    & 2.4324 & 44760 \\
\end{xltabular}
\end{ThreePartTable}

\end{document}